\documentclass[pdftex,twocolumn,epjc3]{svjour3}
\pdfoutput=1
\RequirePackage[T1]{fontenc}
\smartqed
\RequirePackage{graphicx}
\RequirePackage{amsmath,amssymb,bm}
\RequirePackage{newtxtext,newtxmath}
\RequirePackage{booktabs}
\RequirePackage[final]{microtype}
\RequirePackage{placeins}
\RequirePackage{cuted}
\RequirePackage{xcolor}
\RequirePackage[numbers,sort&compress]{natbib}
\RequirePackage[colorlinks=true,citecolor=blue!55!black,urlcolor=blue!55!black,linkcolor=blue!55!black]{hyperref}
\graphicspath{{figures/}}
\newcommand{\DM}{\mathrm{DM}}
\newcommand{\Glam}{\mathrm{G}\lambda}

\newcommand{\dreq}{d_{95,75}}
\journalname{Eur. Phys. J. C}

\begin{document}
\title{Dark matter deformations of photon-ring echoes in horizon-scale interferometry}
\author{Mohsen Fathi\thanksref{e1,addr1}}
\thankstext{e1}{e-mail: mohsen.fathi@ucentral.cl}
\institute{Centro de Investigaci\'on en Ciencias del Espacio y F\'isica Te\'orica (CICEF), Universidad Central de Chile, La Serena 1710164, Chile\label{addr1}}
\date{}
\maketitle

\begin{abstract}
Photon-ring echoes can probe the delayed light paths near rotating black holes, even when their main ring scales agree. We study an effective rotating geometry with a strong dark matter spike and compare it with a separately ring-matched Kerr control. At $a/M_B=0.8$, $i=70^\circ$, and $M_H/M_B=10$, the remaining angular critical-parameter fingerprint is $1.49\%$. The first two higher image orders have median delays of $13.33M$ and $33.06M$, with spike--Kerr shifts of $1.192M$ and $0.378M$. Finite-order ray transport, slow-light synchrotron images, and complex very long baseline interferometry (VLBI) visibilities show distinct delay and interference patterns. Direct and higher-order images contribute differently to the complex visibility signal. With the same emitting source, the model difference survives controlled station noise. Independent synthetic-source tests show that the signal depends strongly on source variability: EHT-like held-out closures give an area under the receiver operating characteristic curve (AUC) of $0.658$ (95\% interval $0.497$--$0.807$), while two source-conditioned tests give AUC values close to $0.5$. Published general relativistic magnetohydrodynamic (GRMHD) light curves of Sgr A* illustrate how electron physics changes variability on the relevant time scales. Together these results connect environmental photon transport to interferometric observables and quantify the importance of source information for future echo studies.
\end{abstract}

\section{Introduction}\label{sec:intro}

The Event Horizon Telescope (EHT) has made horizon-scale black hole structure observable. The first M87* campaign found a ring-like source with several independent calibration and imaging methods \cite{EHT2019M87I,EHT2019M87II,EHT2019M87III,EHT2019M87IV,EHT2019M87V,EHT2019M87VI}. Polarimetric observations later added information about the magnetic field near the horizon \cite{EHT2021M87VII,EHT2021M87VIII}. The Sgr A* campaign showed a second horizon-scale ring and also made rapid variability and scattering central parts of the analysis \cite{EHT2022SgrAI,EHT2022SgrAII,EHT2022SgrAIII,EHT2022SgrAIV,EHT2022SgrAV,EHT2022SgrAVI}. Multi-epoch M87* work and the EHT mid-range plan now put time dependence and stronger spacetime tests among the main goals \cite{EHT2024PersistentI,EHT2025PersistentII,EHTMidRange2024}.

A horizon-scale image contains more information than one ring diameter. Photons close to the critical curve can orbit the black hole before reaching the observer. They form narrow higher-order images whose widths, positions, delays, rotations, and flux ratios are linked to near-critical photon motion \cite{GrallaHolzWald2019,GrallaLupsasca2020,GrallaLupsascaObservable2020,GrallaLupsascaMarrone2020,HadarKapec2022}. Their long-baseline visibility structure has been studied for thin rings, extended flows, and finite-width photon rings \cite{Johnson2020Universal,Vincent2022,Paugnat2022,CardenasLupsasca2023,Jia2024,Farah2026PhotonRingSpin}. Critical parameters can also be used as spacetime diagnostics beyond Kerr \cite{Walia2025,Wan2026KBR,BenAchour2025}.

A variable source adds a time-domain signal, including the delayed lensing echoes that motivated this study. Earlier work discussed black hole glimmer and coherent light echoes \cite{Wong2021Glimmer,Chesler2021Echo}, and later studies developed photon-ring autocorrelations and spectro-temporal correlations \cite{Hadar2021,HadarHarikeshChelouche2023}. Recent calculations show that lensing delays can appear in VLBI observables and time-resolved images even when the total-flux autocorrelation is weak \cite{Wong2024Echo,CardenasGammieLupsasca2024,Zhang2025Echo,Bezdekova2026}. We use ``echo'' for this delayed higher-order lensing signal.

Dark matter (DM) can modify this strong-field transport. Navarro--Frenk--White (NFW), Einasto, Burkert, and Hernquist profiles are standard reference models \cite{NFW1997,Hernquist1990,Burkert1995,Merritt2006Einasto}, while a central black hole can create a much steeper inner spike \cite{GondoloSilk1999,Sadeghian2013,Ferrer2017}. Many studies have considered shadows, lensing, or orbital effects of black holes in DM halos or spikes \cite{Hou2018SgrDM,Hou2018PFDM,XuHouGongWang2018,Nampalliwar2021,XuWangTang2021,KonoplyaZhidenko2022}. Rotating environmental solutions and recent rotating-DM image studies bring this question closer to horizon-scale observations \cite{FernandesCardoso2025,Yue2026,Fathi2026Ray,Huang2026PFDM,AraujoFilho2026Hernquist,Feng2026RotatingDMShadow}. Related strong-field calculations also show that an environmental correction can be much smaller after a careful control match than in a raw shadow-size comparison \cite{FathiAhmed2026,Fathi2026DarkHide}.

Recent multi-frequency closure analyses considered higher-order rings in spherical DM environments \cite{PradhanGhaderiZeyauddin2026}. We extend the problem to a rotating spike geometry and follow its matched-Kerr differences from photon delays to variable emission, complex visibilities, and independent-source tests. This time-dependent analysis builds on earlier static images \cite{Fathi2026Ray}.

The emitting source can strengthen or mask a small metric signature. Calibration, sparse Fourier coverage, variability, and image assumptions already affect horizon-scale inference \cite{EHT2019M87III,EHT2019M87IV,Lu2016}. General relativistic magnetohydrodynamic (GRMHD) models show a wide range of emitting structures and variability states \cite{Porth2019,EHT2019M87V,EHT2022SgrAV}. We also compare the source-variability scales with published GRMHD-based Kerr light curves for Sgr A* \cite{Salas2025GRMHD}. Instrument noise and source fluctuations matter for photon-ring inference \cite{CardenasKeebleLupsasca2024}, and visibility phase can carry information that is different from amplitude \cite{GutierrezLara2026}. The analysis therefore follows two linked questions: how the metric changes the signal after ring matching, and how that signal behaves as source conditions vary.

We first use a shared-source control to isolate the metric response, then test its robustness with independent emission realizations. Our $70^\circ$ benchmark highlights the strong-field effect, while $17^\circ$ gives a lower-inclination comparison motivated by M87* viewing models \cite{Mertens2016M87}. We use the M87* mass and distance to express the angular and time scales.

The calculation follows the hierarchy
\begin{equation}
\begin{aligned}
 g_{\mu\nu}^{\rm eff} &\longrightarrow \{\gamma,\delta,\tau\} \\
 &\longrightarrow T_n(t,x,y,\nu)
 \longrightarrow V(u,v,t,\nu) \longrightarrow {\cal L}.
\end{aligned}
\label{eq:ladder}
\end{equation}
Here $\gamma$, $\delta$, and $\tau$ describe local instability, azimuthal transport, and time delay near the critical photon structure. $T_n$ is the finite-order transfer, $V$ is the complex visibility, and ${\cal L}$ denotes the likelihood used where a probabilistic model is specified; other tests use empirical scores. Sections~\ref{sec:geometry}--\ref{sec:positive} develop the geometry, radiative transfer, and delays. Sections~\ref{sec:visorders}--\ref{sec:req} examine interferometric signals, source robustness, and observing requirements. Section~\ref{sec:externalgrmhd} compares the source-variability scales with published GRMHD light curves; the discussion and conclusions follow.

\section{Matched geometries and finite-order photon transfer}\label{sec:geometry}

\subsection{Geometry families and the matched-Kerr control}

We use three near-critical quantities. The parameter $\gamma$ measures radial instability and sets the narrowing of successive near-critical images. The parameter $\delta$ measures their azimuthal advance, and $\tau$ measures the time delay. We compare them only after matching the leading critical-curve scale.

For each DM target, we build a Kerr control with the same leading critical-curve scale. We also match the main aspect ratio when needed. This leading-scale match provides a common reference for comparing angular and time-dependent transport.

The model set contains vacuum Kerr, smooth Einasto and cored-NFW examples, a density-integrated Hernquist cusp, and a central spike. The smoother profiles are low-signal controls. The spike is the main benchmark because it can extend into the photon region. Earlier halo and spike metrics show that the effect depends strongly on the inner profile and on the way the black hole and environment are combined \cite{Hou2018SgrDM,Hou2018PFDM,XuHouGongWang2018,Nampalliwar2021,XuWangTang2021,KonoplyaZhidenko2022}. Relativistic spike calculations also show that the near-black-hole distribution cannot always be treated as a Newtonian profile \cite{Sadeghian2013,Ferrer2017}. The model set spans smooth, cusped, and spike-like environmental profiles, with the spike providing the main strong-field benchmark.

For the strong spike we use the rotating construction of Yue, Zhao, and Qian (YZQ) \cite{Yue2026}. We define
\begin{equation}
 A=r^2+a^2,\qquad f(r)=2q(r)r,\qquad B(r)=2m(r)r,
\end{equation}
and
\begin{equation}
 \Delta_f=A-f,\qquad \Delta_B=A-B.
\end{equation}
For a photon of conserved energy $E$, axial angular momentum $L_z$, and Carter-like separation constant $Q$, we write $\xi=L_z/E$ and $\mathcal I^2=(\xi-a)^2+Q/E^2$. The separated radial potential for null geodesics is
\begin{equation}
 R(r)=\frac{\Delta_B}{\Delta_f}
 \left[(A-a\xi)^2-\Delta_f\mathcal I^2\right].
 \label{eq:radialpotential}
\end{equation}
The spherical-orbit conditions determine the critical impact parameters through $q(r)$. The second radial function changes the local instability. In our notation,
\begin{equation}
 \gamma_{q,m}=\left(\frac{\Delta_B}{\Delta_f}\right)^{1/2}\gamma_q.
 \label{eq:gammaqm}
\end{equation}
The baseline mass function is
\begin{equation}
 m(r)=M_B+M_H\left(\frac{r}{r_{\rm SP}+r}\right)^\alpha
 \left(1-\frac{r_{\rm ISCO}}{r}\right)^2
 \label{eq:massfunction}
\end{equation}
for $r>r_{\rm ISCO}$, and $m=M_B$ inside. We obtain $q$ from
\begin{equation}
 q'=\frac{m-q}{2m-r},\qquad q(r_{\rm ISCO})=M_B.
 \label{eq:qeq}
\end{equation}
The main benchmark has $M_H/M_B=10$, $\alpha=0.5$, $r_{\rm SP}/M_B=10^4$, and $a/M_B=0.8$. We use this deliberately strong benchmark to measure the response of near-critical photon transport. At the nominal M87* mass, $r_{\rm SP}\simeq3.1$ pc and the profile parameter corresponds to an enclosed excess of $7.07M_B$ at that radius (about $4.6\times10^{10}M_\odot$ for $M_B=6.5\times10^9M_\odot$). The extended mass sets a deliberately strong benchmark for the response of near-critical photon transport. Near the photon region, the excess is much smaller. For $r_{\rm ISCO}=2.907M_B$, the local excess $m(r)-M_B$ is only $1.68\times10^{-4}M_B$ at $r=3M_B$, $5.38\times10^{-3}M_B$ at $3.5M_B$, and $1.49\times10^{-2}M_B$ at $4M_B$. At $i=70^\circ$, the spherical orbits that form the displayed critical curve span $1.842\le r/M_B\le3.776$. Thus, only the outer part of the critical-orbit family reaches the region where the spike mass starts to grow. Figure~\ref{fig:crit}(d) shows this directly.

The YZQ geometry supplies the effective rotating background for our photon calculations. Its separability, Kerr limit, horizon, and $q$--$m$ relations are checked in \ref{app:yueaudit}, with a separate static-seed Einstein-equation check in \ref{app:staticseed}. We use this geometry for its defined photon-transport predictions; the microphysical identification of a rotating DM source is a distinct dynamical question \cite{FernandesCardoso2025}.

\subsection{Critical parameters and an angular fingerprint}

For every screen angle $\psi$, we calculate the critical-curve transport parameters and subtract the separately matched Kerr control. We summarize the remaining angular structure with
\begin{equation}
\begin{aligned}
 {\cal F}_{\rm crit} = \Big\langle \Big|&\bm{X}_{\rm target}(\psi)-\bm{X}_{\rm Kerr}(\psi)\\
 &-\left\langle\bm{X}_{\rm target}-\bm{X}_{\rm Kerr}\right\rangle_\psi
 \Big|^2 \Big\rangle_\psi^{1/2}.
\end{aligned}
 \label{eq:fingerprint}
\end{equation}
Here $\bm X=(\gamma/\pi,\tau/(\sqrt{27}\,\pi M_B),\delta/\pi)$. Removing the screen-angle mean isolates the nonuniform part of the transport deformation under the stated leading-scale match. Related work on photon-ring shape and critical parameters shows why angle-dependent information can contain more than one global size \cite{GrallaLupsascaObservable2020,GrallaLupsascaMarrone2020,Paugnat2022,Walia2025,Wan2026KBR}.

For representative smooth or cusped models at $a/M=0.7$ and $i=70^\circ$, the legacy-amplitude fingerprints are $0.092\%$ for cNFW, $0.090\%$ for Hernquist, and $0.084\%$ for Einasto. With fixed-ADM matching, they fall to $0.0036\%$, $0.0128\%$, and $0.0046\%$. Here ``legacy'' denotes the earlier smooth-profile amplitude normalization and ``fixed-ADM'' holds the asymptotic mass fixed. Both comparisons use $a/M=0.7$; the spike uses $a/M_B=0.8$. The smaller fixed-ADM fingerprints illustrate the importance of the matching convention.

The spike is stronger. At $a/M_B=0.8$, the Kerr-matched fingerprints are $0.206\%$, $0.994\%$, and $1.494\%$ for $i=17^\circ$, $45^\circ$, and $70^\circ$. The three inclinations have different roles: $70^\circ$ is the high-inclination stress test, $17^\circ$ is the low-inclination check, and $45^\circ$ only shows the geometrical trend. The $17^\circ$ geometrical fingerprint is about seven times smaller than at $70^\circ$. The visibility and closure differences need not follow the same ratio because radiative transport and source variability are nonlinear.

\begin{figure*}[t]
 \centering
 \includegraphics[width=0.94\textwidth]{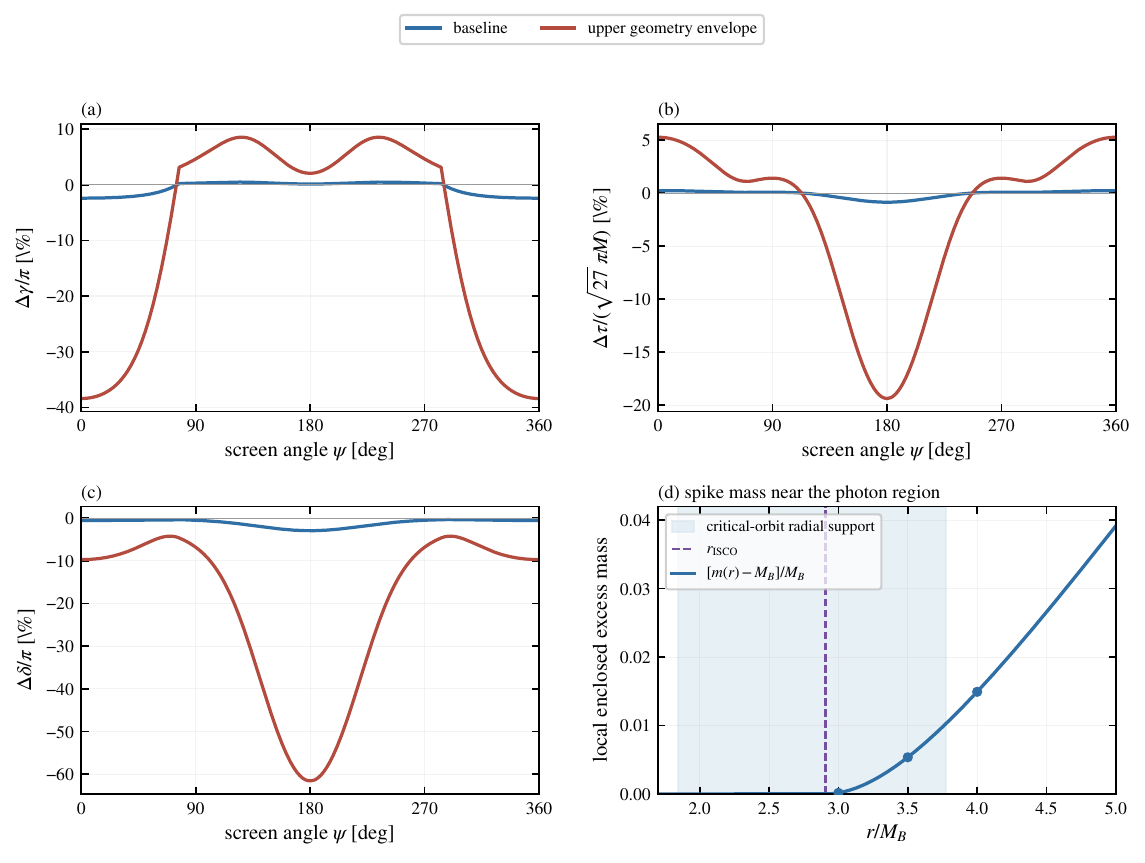}
 \caption{Strong-spike geometry at $a/M_B=0.8$ and $i=70^\circ$. Panels (a)--(c) show the critical-parameter residuals relative to the separately ring-matched Kerr control. Blue curves are the $M_H/M_B=10$ benchmark; red curves show the $M_H/M_B=100$ geometry envelope. The horizontal angle $\psi$ labels location around the observer-screen critical curve. Panel (d) shows the local enclosed excess mass of the baseline spike. The shaded band is the radial support of the spherical orbits that generate the displayed critical curve, and the dashed line is $r_{\rm ISCO}=2.907M_B$. The rising spike mass overlaps the outer part of the critical-orbit family.}
 \label{fig:crit}
\end{figure*}

Figure~\ref{fig:crit} shows the angular structure. The response is not uniform around the screen. Some parts of the photon shell remain inside the Kerr-like region, while other parts reach the spike-active region. We also show an $M_H/M_B=100$ envelope only as a geometry-level reference.

\subsection{Finite image orders}

We build source-independent transfer kernels for image orders $n=0,1,2$. An adaptive mesh integrates the finite-order bands on both sides of the exact separatrix, following the same basic idea as adaptive photon-ring ray tracing \cite{AART2023}. The cell that crosses the separatrix is treated as an improper-integral limit, so the infinite-dwell critical orbit never receives an ordinary pixel weight. The adaptive gate passes for the main spike, the low-inclination spike, a $C^2$-smoothed spike onset, and the Hernquist fixed-ADM control.

One echo order contains several pieces of information. We use
\begin{equation}
 \Delta\tau_n,\qquad \Delta\delta_n,\qquad
 \frac{{\cal R}_{n}^{\rm target}}{{\cal R}_{n}^{\rm Kerr}},
 \label{eq:finiteobservables}
\end{equation}
for the delay, rotation, and finite-order response. Each quantity is first measured relative to the direct image and then compared with matched Kerr. Rotation is measured with a non-axisymmetric source statistic, because an axisymmetric ring has no useful orientation.

\begin{figure*}[t]
 \centering
 \includegraphics[width=0.94\textwidth]{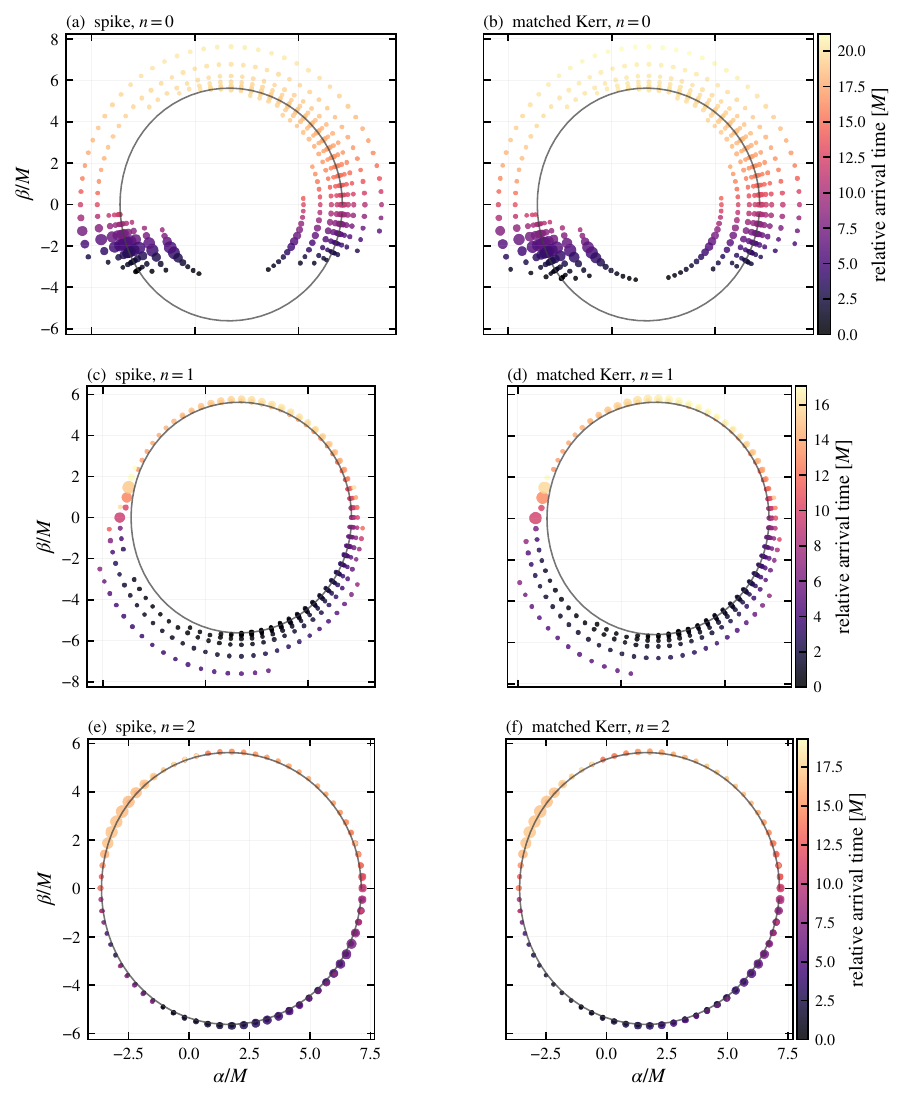}
 \caption{Finite-order transfer for the strong spike and its separately ring-matched Kerr control. The rows show image orders $n=0,1,2$. The left and right columns show the target and the matched control. After matching the main ring scale, the transfer maps retain distinct delay, rotation, redshift, and lensing-weight patterns.}
 \label{fig:transfer}
\end{figure*}

Figure~\ref{fig:transfer} compares the order-resolved transfer maps before the radiative source is introduced. For the $a/M_B=0.8$, $i=70^\circ$ spike, the order-1 response ratio is $1.328$ and the differential rotation is $1.69^\circ$. For order 2, the values are $1.342$ and $2.70^\circ$. The absolute delay residual depends on how the leading ring match is interpreted. If the scale change is absorbed as a mass-distance rescaling, the sign can differ from a convention with an external mass prior. For this reason, we keep timing and angular scale separate in the later observational discussion.

\section{Radiative source and visibility model}\label{sec:radiative}

\subsection{Slow-light synchrotron source}

We reuse the converged geometric kernels and place a thin equatorial thermal synchrotron source on the transfer map. The emissivity and absorption follow standard relativistic synchrotron prescriptions \cite{Pandya2016}. Source fluctuations are an advected anisotropic Gaussian random field, based on the stochastic disk model of Ref.~\cite{LeeGammie2021}. This simple source lets us vary the main fluctuation and transfer properties in a controlled way. GRMHD simulations provide complementary information on magnetized flows and their variability \cite{Porth2019,EHT2019M87V,EHT2021M87VIII,EHT2022SgrAV}.

Each image-order crossing uses its retarded emission time. Fast light is kept as a control. Slow light matters when the source changes on a time scale comparable with the lensing delay \cite{RojasPaternina2026}. In the fiducial M87*-scaled spike movie, it changes the total flux by a fractional RMS of about $19\%$, with larger changes for faster or stronger variability.

\begin{figure*}[t]
 \centering
 \includegraphics[width=0.94\textwidth]{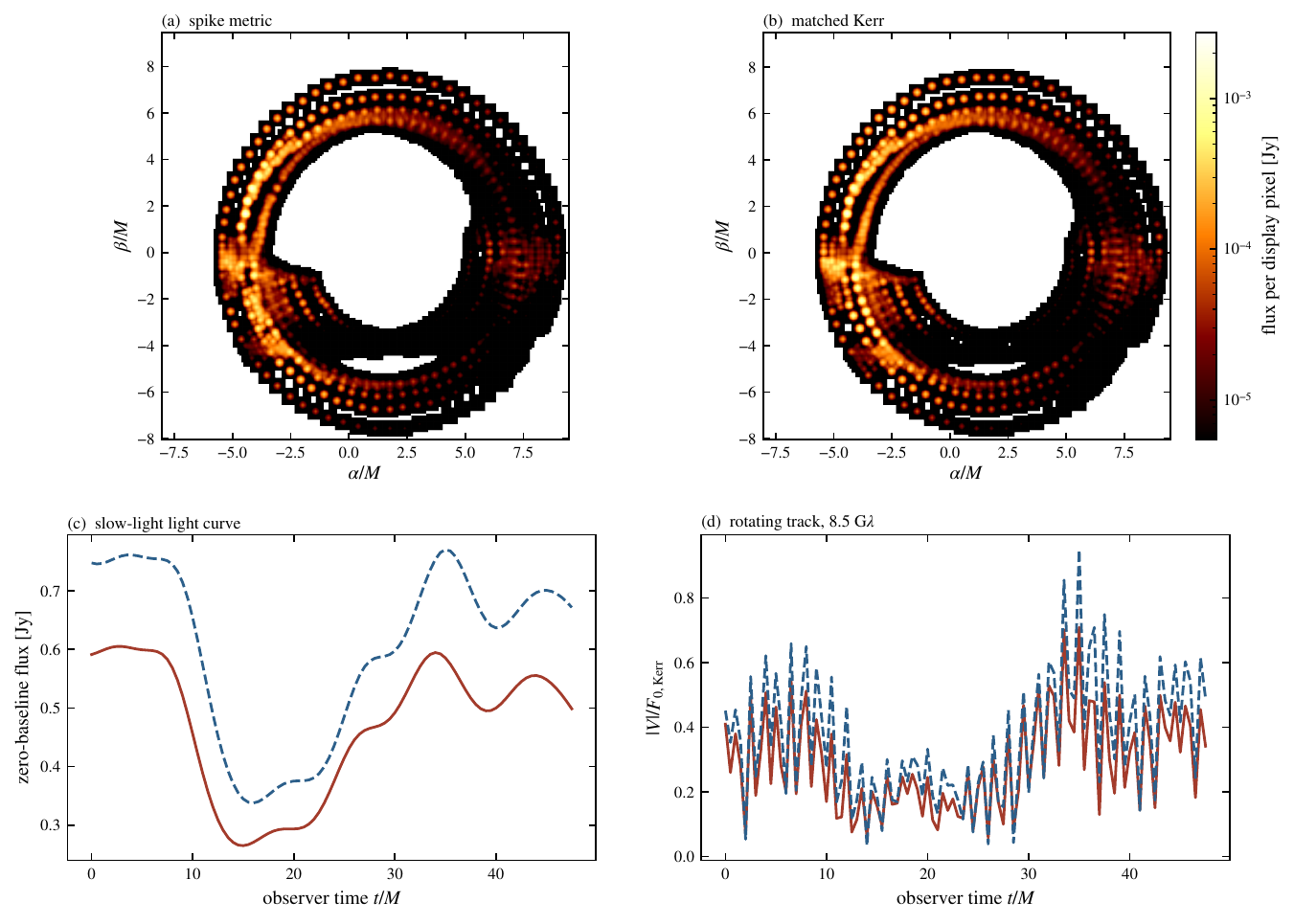}
 \caption{One representative radiative realization. Panels (a) and (b) show the target and matched-Kerr images. In panels (c) and (d), solid red is the spike case and dashed blue is matched Kerr. The line-style key is displayed inside panel (c). Panel (c) shows the slow-light total flux, and panel (d) shows a long-baseline visibility track. Higher image orders use delayed source states. Their detailed amplitude is also changed by redshift, absorption, and lensing weight.}
 \label{fig:movie}
\end{figure*}

Figure~\ref{fig:movie} displays an example ray-traced target/Kerr image pair and its time-dependent signals. We pair the same archived stochastic source realization across both geometries and use one flux normalization determined from matched Kerr. The two ray-traced outputs retain their geometry-dependent transport and redshift. The main frequency is 230 GHz, with 345 GHz used as a high-frequency consistency check. For the same latent source, the target-to-Kerr mean-flux ratio is $0.783$ at 230 GHz and $0.775$ at 345 GHz. The RMS target--Kerr difference of the total correlation map is $0.0122$ and $0.0116$. The EHT-like 345-GHz calculation also passes the closure and fine/coarse convergence gates, with worst 95th-percentile normalized visibility error $0.0206$. The two frequencies sample the same underlying source realization. The check is relevant to current EHT and future high-frequency plans \cite{EHTMidRange2024,BHEX2024}.

\subsection{Complex visibilities and closures}

For adaptive screen nodes $(x_j,y_j)$ with weights $w_j$ and intensities $I_j(t,\nu)$, we calculate the complex visibility directly,
\begin{equation}
 V(u,v,t,\nu)=\sum_j w_j I_j(t,\nu)
 \exp\left[-2\pi i(ux_j+vy_j)\right].
 \label{eq:visibility}
\end{equation}
This direct sum avoids raster interpolation across very narrow photon-ring bands. Fine/coarse adaptive-tree comparisons define the visibility-convergence gate. We keep the 230-GHz ground-baseline range where the predefined p95 complex-visibility difference is below $3\%$.

\begin{figure*}[t]
 \centering
 \includegraphics[width=0.94\textwidth]{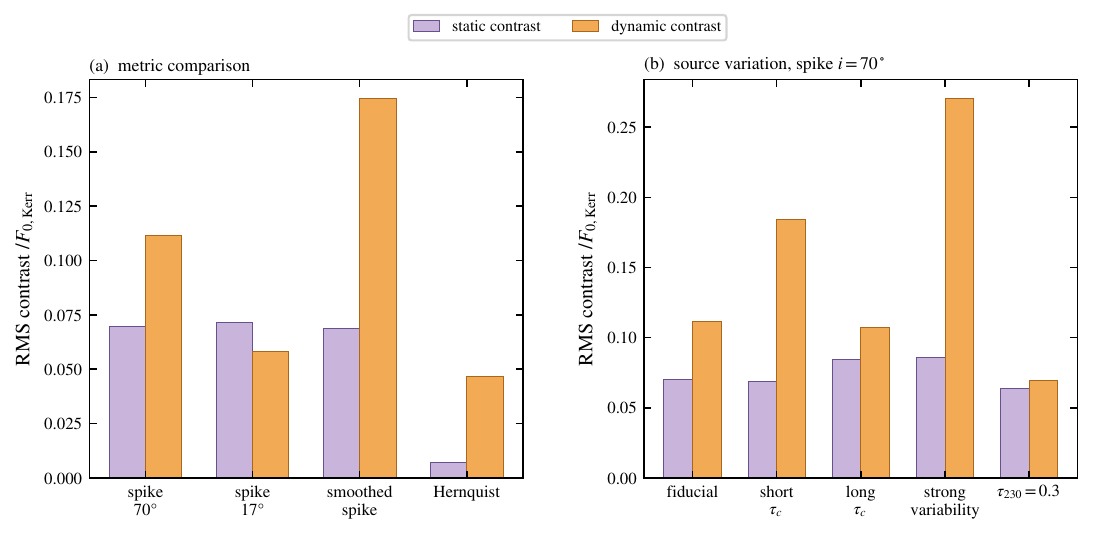}
 \caption{Static and dynamic visibility contrast in the controlled source model. Panel (a) compares several geometries. Panel (b) changes source properties for the main spike. Panel (b) uses reference optical depth $\tau_{230}=0.3$ at $230$ GHz. The dynamic contrast is often larger than the static contrast and varies with the emission model.}
 \label{fig:contrast}
\end{figure*}

The main radiative quantity is
\begin{equation}
\begin{aligned}
 \Delta V_{\DM}(u,v,t,\nu) ={}& V_{\rm target}(u,v,t,\nu)\\
 &-V_{\rm matched\ Kerr}(u,v,t,\nu).
\end{aligned}
 \label{eq:deltav}
\end{equation}
Figure~\ref{fig:contrast} compares the static and dynamic visibility contrasts across model and source choices. We separate a repeated-source static part from the extra time-dependent part. At 230 GHz, the high-inclination spike has static contrast $0.0698F_0$ and dynamic contrast $0.1117F_0$, where $F_0$ is the matched-Kerr zero-baseline flux. The contrast combines photon transport, gravitational redshift, and emission weighting. An image-order check on the archived movie, including independent mean-flux scaling, is given in \ref{app:ordercheck}. Across the bounded source and absorption tests the contrast ranges from about $0.069$ to $0.271$.

The Hernquist arm shows the same source dependence at lower amplitude: $0.007F_0$ static and $0.047F_0$ dynamic. The source dependence motivates the independent-source tests below. The model response is distributed over baseline, lag, angle, and frequency, as also found in wider photon-ring studies \cite{Vincent2022,Paugnat2022,Jia2024,GutierrezLara2026}.

\section{Delay structure in the forward model}\label{sec:positive}

We compare the delays from the photon-transfer calculation with the correlations in the simulated emission.

For the main spike, the weighted order-1 delay distribution has a 16th--84th percentile interval of $10.95M$--$19.38M$ and median $13.33M$. The corresponding order-2 interval is $28.43M$--$38.88M$, with median $33.06M$. With the M87* time scale used in our controlled forecast, these medians are about 4.94 and 12.24 days. A longer diagnostic movie shows order-tagged visibility-correlation peaks at $12.0M$ and $28.5M$. Both lie inside the broad geometric delay supports. The unresolved total-flux autocorrelation has no local maximum at these delays.

\begin{figure*}[t]
 \centering
 \includegraphics[width=0.94\textwidth]{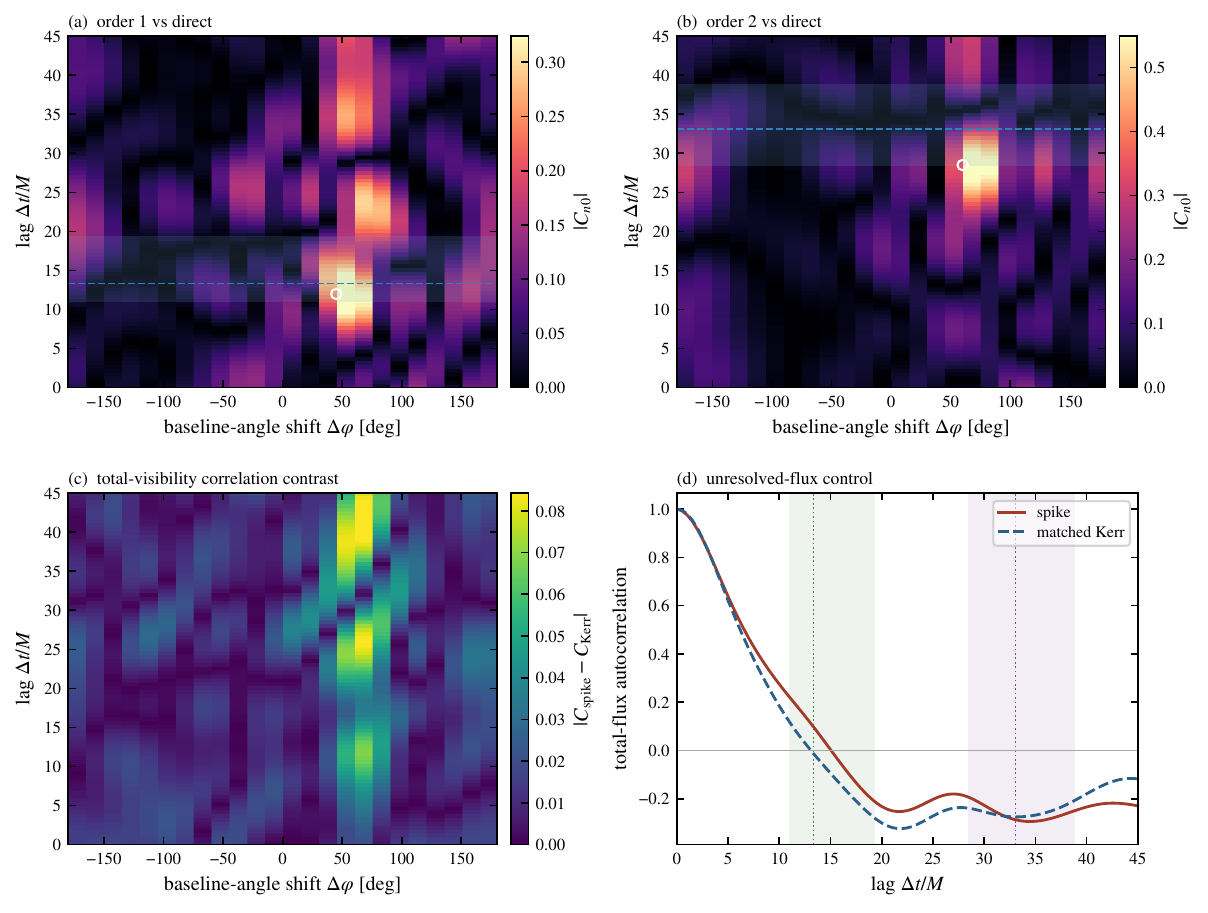}
 \caption{Delay-anchored correlations for the M87*-scaled strong-spike benchmark. Panels (a) and (b) show order-1/direct and order-2/direct visibility correlations. The dashed lines mark the geometry-weighted delay medians and the white circles mark radiative peak candidates. Panel (c) shows the spike-minus-Kerr contrast in total-visibility correlation. Panel (d) shows the unresolved total-flux autocorrelation. Resolved structure appears inside the geometric delay supports even though the total light curve has no clear second peak there. The white markers identify local radiative maxima within the geometric delay windows.}
 \label{fig:echo-corr}
\end{figure*}

This behavior is expected from recent echo studies. A simple light curve can lose the secondary peak because many source regions contribute at different phases \cite{CardenasGammieLupsasca2024}. More resolved observables can keep lensing information \cite{Hadar2021,HadarHarikeshChelouche2023,Wong2024Echo,Bezdekova2026,Zhang2025Echo}. Figure~\ref{fig:echo-corr} shows the same effect in our DM-minus-Kerr problem. The baseline-resolved maps have localized structure close to the geometric delay windows, while the unresolved total-flux control is smooth through the same region.
The candidate peaks connect the geometric delay windows with the radiative signal; nearby aliases and source coherence affect their prominence. The inference below uses the full visibility or closure trajectory.
The effect also survives a smoother spike edge. A $C^2$-smoothed spike onset keeps degree-scale differential rotations and order-unity response-ratio changes, although some delay residuals change. The $i=17^\circ$ case is smaller but not zero. These controls show that onset shape and viewing orientation affect the strength of the environmental response.

\section{Higher-order visibility signatures}\label{sec:visorders}

The finite-order ray transfer also changes the complex Fourier data. We now examine two diagnostics of that change using the archived, resolved-order visibilities. They show how the direct and higher-order images combine in one controlled source realization.

\subsection{Cancellation in complex visibilities}\label{sec:interference}

\begin{figure*}[t]
 \centering
 \includegraphics[width=0.98\textwidth]{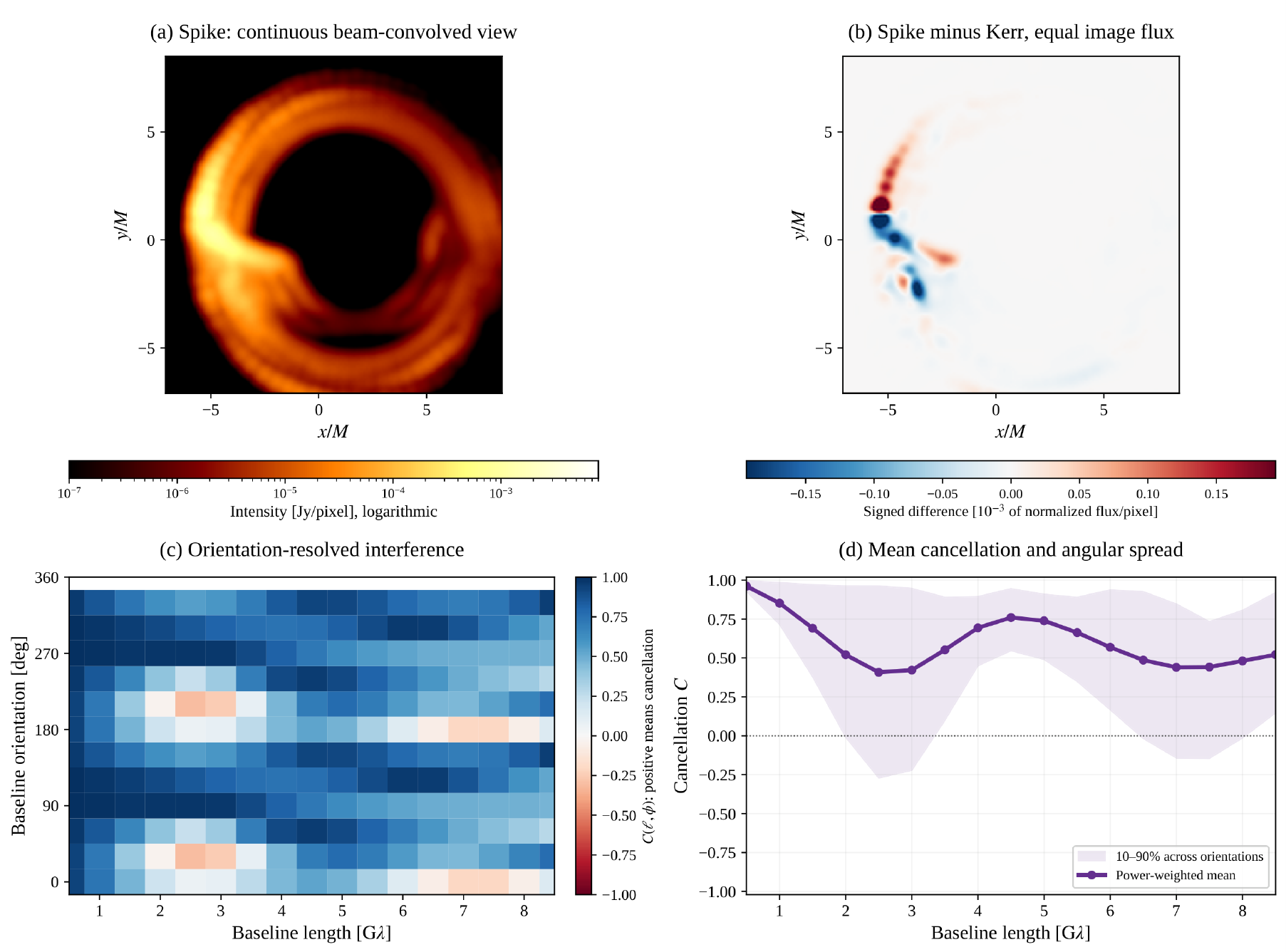}
 \caption{Emission and image-order interference at $230$ GHz. Panel (a) shows the spike at $t=36M$ with Gaussian display smoothing ($\sigma=0.25M$). Panel (b) shows the flux-normalized spike-minus-Kerr image residual after the same smoothing and regridding. Panel (c) gives $C(\ell,\phi)$ for 12 baseline directions, and (d) gives the power-weighted mean with its 10--90\% angular range. All visibility calculations use the original, unsmoothed complex data. The emission maps use the semianalytic synchrotron source described in Sec.~\ref{sec:radiative}.}
 \label{fig:imagetransport}
\end{figure*}

The complex visibility is a sum over image orders. We use the same archived source realization and flux factor $g$ as in \ref{app:ordercheck}. We form $D_0=gV_{0,{\rm sp}}-V_{0,K}$ and $D_h=g(V_{1,{\rm sp}}+V_{2,{\rm sp}})-(V_{1,K}+V_{2,K})$. For each static baseline length we average over 12 directions and define
\begin{equation}
 C=-\frac{2\,\mathrm{Re}\langle D_0 D_h^*\rangle}
 {\langle |D_0|^2\rangle+\langle |D_h|^2\rangle}.
 \label{eq:ordercancel}
\end{equation}
A positive $C$ means that direct and higher-order differences partly cancel. We find $C=0.409$, $0.740$, and $0.481$ at $2.5$, $5$, and $8\,\mathrm{G}\lambda$. Figure~\ref{fig:imagetransport} shows a beam-convolved display of one ray-traced image, its equal-flux difference from Kerr, and the angular dependence of $C$. Only the displayed image maps are smoothed; all visibility results use the original numerical arrays. Across the five saved images, half the pixelwise $L^1$ distance between separately flux-normalized rasters is $0.231$--$0.248$ after regridding. The image and Fourier diagnostics use the same five archived movie frames.

\subsection{Visibility spectra of direct and higher-order images}\label{sec:ordervlbi}

We also use the archived static complex visibilities to examine how the first higher-order image changes with baseline length. At $230$ GHz the stored grid contains 27 baseline lengths from $0$ to $20\,\mathrm{G}\lambda$ and 12 directions. We evaluate $|V_0|$, $|V_1|$, and $|V_0+V_1|$ from these complex values, not from image amplitudes. In the $0$--$10\,\mathrm{G}\lambda$ part of this grid, $|V_1|<|V_0|$ at every sampled direction and length. The largest ratio is $0.537$. Thus, in this source realization, the first higher-order component does not dominate the direct component on the sampled Earth-scale baselines.

\begin{figure*}[t]
 \centering
 \includegraphics[width=0.98\textwidth]{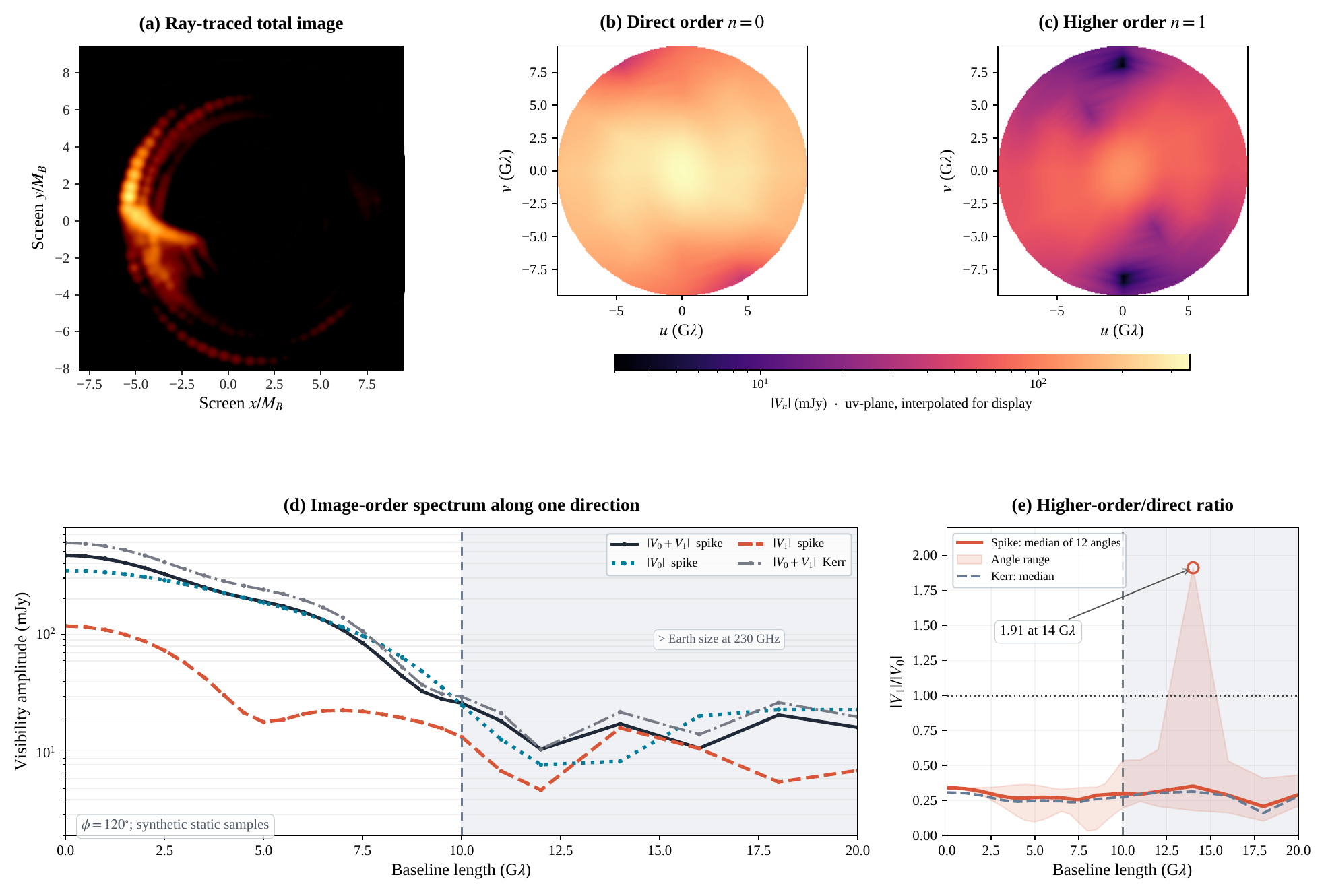}
 \caption{Image-order visibility spectra in the synthetic $230$-GHz spike model. Panel (a) shows one saved total ray-traced image at $t=36M$; a Gaussian width $0.18M_B$ is used for display only. Panels (b,c) show the \emph{Fourier-plane visibility amplitudes} of orders $n=0$ and $n=1$, interpolated between the stored static samples for display. The two panels are plotted in the $(u,v)$ Fourier plane. The color scale in (b,c) is shared. Panel (d) shows the original complex-visibility amplitudes along the sampled direction $120^\circ$, with the matched-Kerr $|V_0+V_1|$ as a control. Panel (e) gives the median $|V_1|/|V_0|$ over 12 directions, its full angular range for the spike, and the Kerr median. The gray region marks synthetic baselines beyond $10\,\mathrm{G}\lambda$. All quantitative curves use original values without image smoothing or interpolation.}
 \label{fig:ordervlbi}
\end{figure*}

At $14\,\mathrm{G}\lambda$ and direction $120^\circ$, the ratio $|V_1|/|V_0|$ is $1.913$ for the spike and $1.344$ for ring-matched Kerr. Here the direct visibility approaches a local minimum, which increases the ratio. These samples extend beyond the approximate Earth-diameter baseline limit at $230$ GHz and illustrate the higher-order signal near a direct-image visibility minimum. Figure~\ref{fig:ordervlbi} displays the model spectra and angular dependence. The comparison is a controlled image-order spectrum for the archived source realization.

\section{Controlled interferometric oracle test}\label{sec:oracle}

\subsection{Station-level forecasts}

We send the same finite-order source realization through three controlled VLBI configurations: an EHT2025-like ground array, a five-site ngEHT-like extension, and a bounded ground-plus-space case. These controlled arrays follow representative ground and future long-baseline layouts \cite{EHT2019M87II,EHTMidRange2024,Issaoun2023,BHEX2024}.

\begin{figure*}[t]
 \centering
 \includegraphics[width=0.94\textwidth]{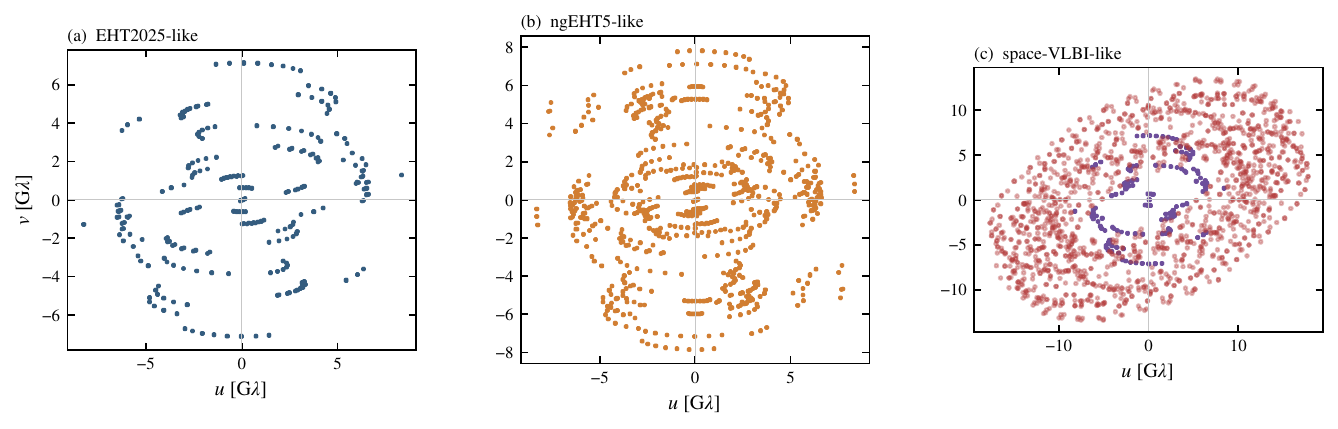}
 \caption{Controlled Fourier coverage for the EHT-like, ngEHT-like, and bounded ground-plus-space configurations. The red and blue points in panel (c) show space--ground and ground--ground baselines, respectively.}
 \label{fig:uv}
\end{figure*}

For every scan, we add station-equivalent system noise, weather and availability draws, baseline flags, and epoch-local correlated complex gains. We then build independent closure phases and log closure amplitudes with shared-baseline covariance. Closure quantities are useful because they reduce sensitivity to station-based gain errors \cite{Chael2018}. EHT calibration and imaging analyses show why such quantities and calibration systematics must be treated carefully \cite{EHT2019M87III,EHT2019M87IV,EHT2022SgrAII,EHT2022SgrAIII}. Figure~\ref{fig:uv} shows the Fourier coverage used here.
All 36 actual-track fine/coarse visibility rows pass the predefined p95 error limit of 0.03 after the allowed one-time refinement of the difficult smoothed-spike and Hernquist controls. This test uses the full time-dependent synthetic station tracks. It is stronger than checking only an azimuthally averaged visibility profile.

\subsection{Shared-source recovery}

The first inference test uses the same stochastic source realization for the target and control libraries. This shared-source comparison tests whether station noise and calibration corruptions preserve the geometry-dependent dynamic signal.

In this oracle check, $\eta=0$ and $\eta=1$ label the Kerr and spike template endpoints. We use 48 instrument and noise realizations for each geometry and array. The coordinate $\eta$ interpolates between the two fixed template endpoints and is used only in this controlled comparison.

\begin{figure*}[t]
 \centering
 \includegraphics[width=0.94\textwidth]{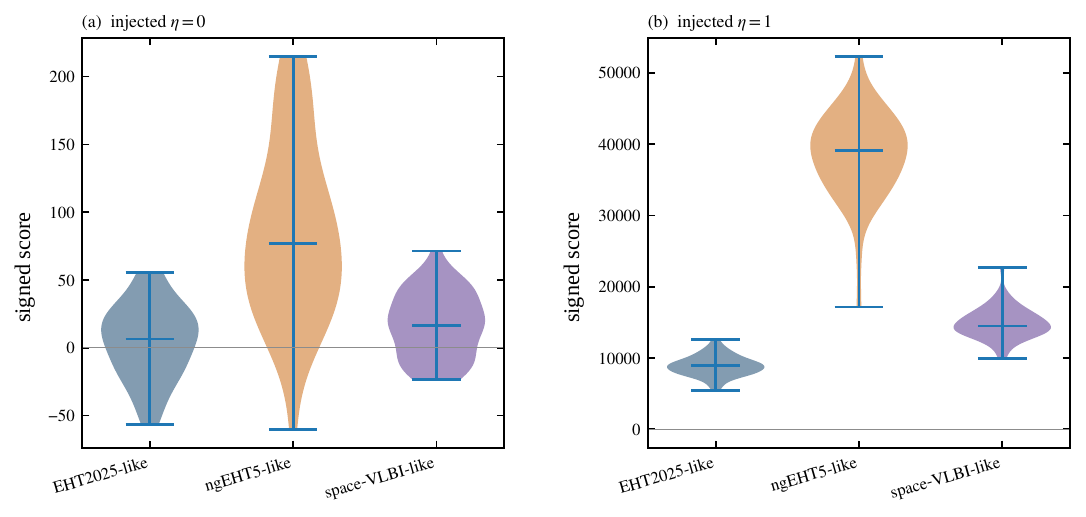}
 \caption{Shared-source oracle test. The vertical axis shows the signed score $\Delta\chi^2=\chi^2_{\rm repeated\ static}-\chi^2_{\rm static+dynamic}$. The violins show its empirical distribution for Kerr-null and spike realizations. The score uses the measured Kerr-null distribution and may take negative values. Its separation shows that the differential dynamic signal survives the simulated noise and station corruptions for a shared source.}
 \label{fig:oracle}
\end{figure*}

We also compare the static-plus-dynamic model with a repeated-static model. We use the signed score $\Delta\chi^2=\chi^2_{\rm repeated\ static}-\chi^2_{\rm static+dynamic}$, calibrated empirically with an array-specific Kerr-null ensemble. The score is calibrated with the empirical Kerr-null distribution, which includes negative values. It is used as a ranking statistic rather than a nested likelihood-ratio significance. The 95th-percentile thresholds are $52.2$, $203.0$, and $50.3$ for the EHT-, ngEHT-, and space-like configurations. The corresponding signed-score medians are $8891$, $39099$, and $14499$. Every injected spike realization exceeds its array-specific empirical threshold. Figure~\ref{fig:oracle} shows the separation obtained in this shared-source comparison.

The shared-source distributions are widely separated in these synthetic tests. We next examine how that separation changes for independent sources.

\section{Source-robust inference}\label{sec:robust}

\subsection{Independent source populations}

We now treat the source realization as unknown. For each geometry, the raw-closure likelihood uses 32 training source realizations and 24 independent held-out injections. The population varies fluctuation amplitude, correlation time, optical depth, radial and azimuthal correlation length, and rotation fraction. Training and held-out injections use disjoint seeds and nuisance vectors. We use wrapped closure phase and log closure amplitude. The covariance is empirical and low rank, with diagonal shrinkage and angular-scale marginalization. For each array, the empirical discrimination score is evaluated on the held-out population. Its decision threshold is the 95th percentile of the corresponding Kerr scores; the false-positive fraction is reported for this same population.

For the EHT-like configuration, the raw population test gives a held-out Kerr false-positive fraction of $0.083$, target power $0.083$, and area under the receiver operating characteristic (ROC) curve (AUC) of $0.658$ (95\% interval $0.497$--$0.807$), and balanced accuracy $0.500$. The ngEHT-like and space-like controls give AUC $0.682$ and $0.623$, with similarly low power. The score distributions overlap broadly across independent sources. \ref{app:metrics} gives the sample counts and 95\% intervals for all three arrays.

The same source index tends to move the score in the same direction across different arrays, with cross-array Spearman coefficients between $0.517$ and $0.756$. We also checked the score against each of the six source-design coordinates. For the EHT-like sample, none gives a stable large correlation in both injected classes: all six have $|\rho_S|\le0.357$ for Kerr and $|\rho_S|\le0.314$ for the spike. A derived realized-field mean is more correlated in the Kerr sample ($\rho_S=-0.496$) but not in the spike sample ($+0.181$). The correlation pattern points to a collective source-realization effect rather than a single dominant design parameter. \ref{app:metrics} lists the six design-coordinate correlations. The additional covariance and temporal-summary controls lead to the same source-dependent behavior. This is consistent with the wider EHT experience that source variability can be a main part of the inference problem, especially for Sgr A* \cite{Lu2016,EHT2022SgrAIII,EHT2022SgrAIV,EHT2022SgrAV}.

\subsection{Conditioning on measured source drivers}

We next condition the long-baseline response on measured source information and run two independent validation tests.

The first test uses total flux with a geometry-specific delayed complex transfer. Development uses 12 Kerr and 12 spike sources and reaches AUC $0.861$. The fixed method is then applied to 16+16 new validation sources. One source in each class fails the predefined response-coverage rule, leaving $15+15$ evaluable sources. Validation falls to AUC $0.507$, with zero target power at the fixed threshold. Fresh validation places the source-conditioned result near chance. \ref{app:metrics} gives the uncertainty intervals.

\begin{figure*}[t]
 \centering
 \includegraphics[width=0.94\textwidth]{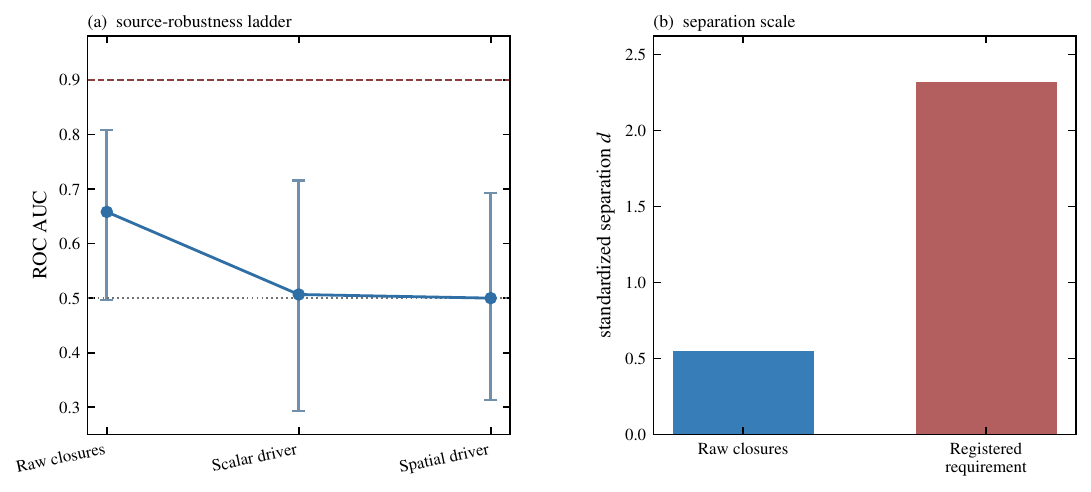}
 \caption{Source robustness and the predefined decision scale. Panel (a) shows the EHT-like ROC area for raw independent-source closures, fresh validation with a total-flux driver, and fresh validation with a compact spatial driver. Error bars are class-stratified 95\% bootstrap intervals (20,000 resamples). The dashed red line is the predefined AUC gate and the dotted gray line is chance. Panel (b) compares the empirical raw-closure standardized separation with the diagnostic $d_{95,75}=2.319$ requirement. Panel (a) compares only independent-source ROC analyses; the shared-source control is shown separately in Fig.~\ref{fig:oracle}.}
 \label{fig:robust}
\end{figure*}

The second test adds nine normalized short-baseline Fourier amplitudes at $0.5$, $1.0$, and $1.5\,\Glam$ in three directions. We combine them with total flux and compress the spatial information into two unsupervised principal components. Development uses $16+16$ sources. Of the 40 new validation sources ($20$ per geometry), $39$ meet the fixed response-coverage criterion, yielding 20 Kerr and 19 spike scores. Fresh validation gives false-positive rate $0.050$, power $1/19=0.0526$, AUC $0.500$, and balanced accuracy $0.501$. The corresponding 95\% intervals are listed in \ref{app:metrics}. Figure~\ref{fig:robust} summarizes the three tests. The three statistics use different source information but address the same practical identifiability question. Their fresh-source results place a quantitative bound on the robustness achieved with the present source models. The reserved confirmatory populations remain unopened.
These validation results motivate the source-control and cadence diagnostics in Sec.~\ref{sec:req}.

\section{Conditional source-control and cadence diagnostics}\label{sec:req}

\subsection{A diagnostic score-separation scale}

Using the same 95th-percentile Kerr threshold, a simple equal-variance Gaussian location model gives a useful design number. To obtain 75\% signal power at this threshold, the standardized mean separation must satisfy
\begin{equation}
 \dreq=\Phi^{-1}(0.95)+\Phi^{-1}(0.75)=2.319.
 \label{eq:dreq}
\end{equation}
At true-negative rate 0.95, the same operating point gives balanced accuracy $0.85$. The condition $\mathrm{AUC}=0.90$ alone would need only $d=1.812$. So Eq.~\eqref{eq:dreq} is the stricter predefined condition.

The raw EHT-like independent-source distribution has pooled $d=0.553$ and robust median/median-absolute-deviation (MAD) value $d=0.537$. This is only about one quarter of the diagnostic requirement. If the mean target--Kerr separation stayed fixed and the total scatter of the test score became smaller, then
\begin{equation}
 \frac{\sigma_{\rm score,new}}{\sigma_{\rm score,old}}
 \lesssim \frac{0.553}{2.319}=0.238.
 \label{eq:scatterreq}
\end{equation}
Under this fixed-score Gaussian approximation the point estimate corresponds to a 76\% reduction of total score scatter, but the uncertainty is large. A 20,000-resample bootstrap gives a residual-scatter range $0.022$--$0.444$. The factor $0.238$ is a conditional score-scatter benchmark that combines source and instrumental contributions.

The two driver tests have different score distributions and are assessed separately.

\subsection{Delay windows and daily gaps}

The transfer function also gives a scheduling condition. In the M87*-scaled setup, the union of the target and matched-Kerr 16th--84th percentile delay windows is
\begin{align}
 n=1:&\quad 4.05\mbox{--}7.38~\mathrm{d},\\
 n=2:&\quad 10.53\mbox{--}14.72~\mathrm{d}.
\end{align}
The order-1 median delays are 4.936 d for the spike and 5.378 d for Kerr. For order 2 they are 12.245 d and 12.385 d. The spike--Kerr median differences are only 10.60 h and 3.36 h.

Our simplified observing schedule uses 51.43-min scan spacing inside a 6-h daily observing block. The scan spacing is short enough to resolve these median differences when the delayed response falls inside the block. The daily gap is the main problem. Broad delay-window pair coverage is already $0.894$ for order 1 and $0.725$ for order 2. With one-scan tolerance, the chosen daily gaps miss both matched-Kerr median-delay phases.

\begin{figure*}[t]
 \centering
 \includegraphics[width=0.94\textwidth]{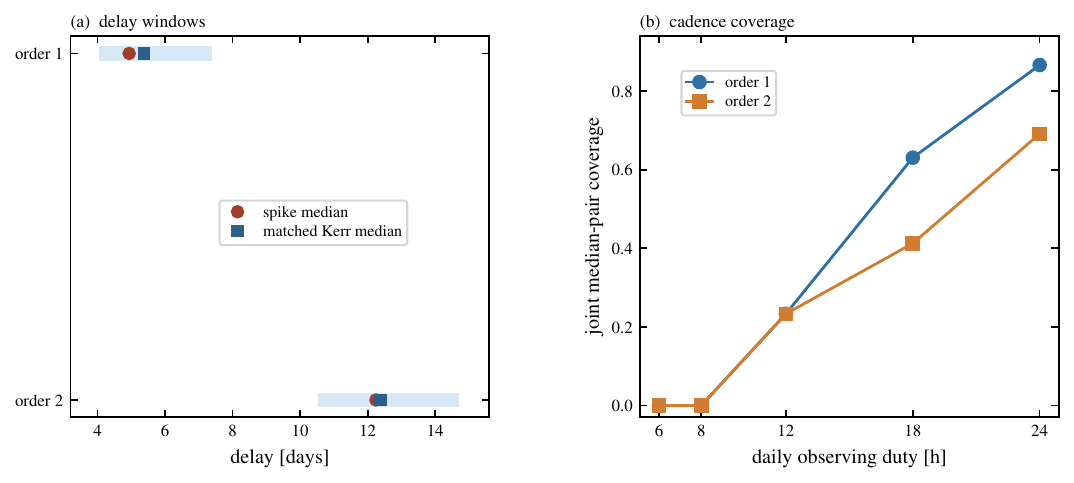}
 \caption{Delay and cadence diagnostic. Panel (a) shows the union delay windows and the spike and Kerr medians for the first two higher image orders. Panel (b) shows a median-pair coverage proxy for a simplified daily observing window as the duty fraction increases. Broad delay-window coverage can coexist with missed median-delay phases.}
 \label{fig:cadence}
\end{figure*}

Figure~\ref{fig:cadence} shows the phase-coverage effect. At an 18-h daily diagnostic duty, the fraction of driver scans with both spike and Kerr median pairs becomes $0.631$ and $0.413$ for orders 1 and 2. At 24 h it becomes $0.866$ and $0.691$. Equivalent phase coverage could come from longitude coverage, coordinated scheduling, or complementary measurements rather than one array observing continuously. These questions are already relevant to future EHT and space-VLBI planning \cite{EHTMidRange2024,BHEX2024}.
The source correlation time gives another duration scale. In the fixed source design, the 16th, median, and 84th percentile correlation times are about 2.71, 5.23, and 7.57 d. Accounting for the longest order-2 delay, campaigns of 40, 60, 75, and 90 d contain about 3.34, 5.98, 7.96, and 9.94 blocks at the source correlation-time 84th percentile. The longer-campaign diagnostics emphasize the role of both duration and delay-phase sampling.

\subsection{Profile hierarchy and observing implications}

\begin{figure*}[t]
 \centering
 \includegraphics[width=0.94\textwidth]{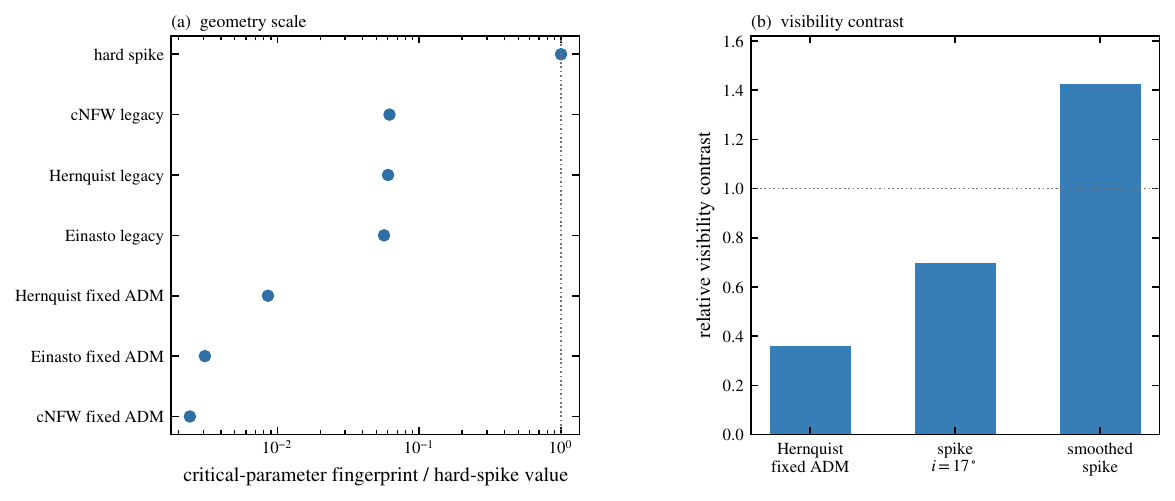}
 \caption{DM profile hierarchy in the controlled model. Panel (a) shows representative critical-parameter fingerprints relative to the hard-spike benchmark. The smooth-profile results use $a/M=0.7$, while the spike normalization uses $a/M_B=0.8$; the ratios show the effect of the model and matching conventions. Panel (b) shows selected shared-source visibility contrasts relative to the same benchmark; the dotted line marks the hard-spike reference value. The geometric and radiative hierarchies differ because the transfer depends on the emitting source.}
 \label{fig:hierarchy}
\end{figure*}

We chose the strong spike because it gives the clearest chance to see an environmental effect near the critical photon region. Figure~\ref{fig:hierarchy} puts the smoother models in context. The representative smooth-profile critical-parameter fingerprints are only a few percent of the hard-spike benchmark and become smaller with fixed-ADM matching.

A first-order geometric scaling suggests that smaller fingerprints need tighter source control, but the radiative calculation changes the ranking. The shared-source total visibility contrast is $0.360$ of the hard-spike value for Hernquist, $0.699$ for the $i=17^\circ$ spike, and $1.425$ for the smoothed spike onset. Source transfer can reorder cases that look simple at the geometry level, consistent with earlier photon-ring studies \cite{Vincent2022,Paugnat2022,Jia2024}.

The two hierarchies clarify how geometry and source transfer jointly determine the relative response across profiles.

\section{Source-variability check with published GRMHD light curves}\label{sec:externalgrmhd}

We examine published Kerr GRMHD-based Sgr A* light curves from the magnetically arrested-disk simulations of Salas et al.\ \cite{Salas2025GRMHD}. The openly released flux arrays were post-processed with the Black Hole Observations in Stationary Spacetimes (BHOSS) ray-tracing code \cite{Salas2025Data}. These simulations provide time-resolved emission from a physical accretion flow and complement our controlled semianalytic model. We use the released flux records to measure source variability and temporal correlations.

\subsection{Time variability of a physical emitting flow}

\begin{figure*}[t]
 \centering
 \includegraphics[width=0.98\textwidth]{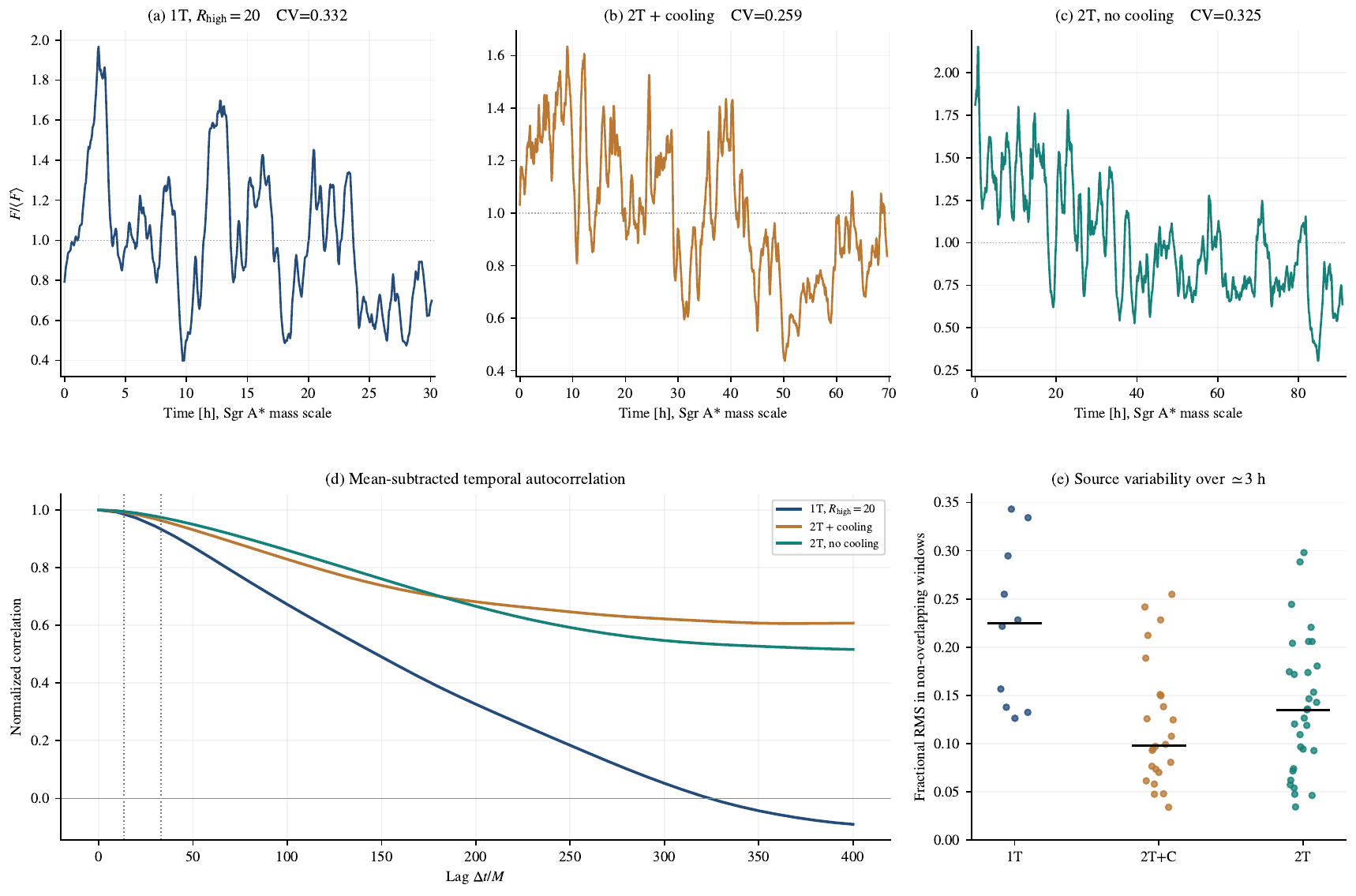}
 \caption{Independent source-variability check using published GRMHD-based synchrotron light curves for Sgr A* at 228 GHz \cite{Salas2025GRMHD,Salas2025Data}. The upper row shows normalized flux time series for one-temperature (1T), two-temperature with cooling, and two-temperature without cooling models. The lower panels show temporal autocorrelations and fractional RMS in nonoverlapping 52-sample (approximately 3-hour) windows, computed from the same released flux records. Dotted markers in panel (d) indicate the two median higher-order delays of the separate rotating-spike photon-transfer calculation. All light-curve panels were derived from the released numerical flux records.}
 \label{fig:grmhdlight}
\end{figure*}

We use the published 228-GHz light curves for the one-temperature (1T), two-temperature with cooling (2T+C), and two-temperature without cooling (2T) models. We normalize each light curve by its own time mean and measure its coefficient of variation, $\sigma_F/\langle F\rangle$. For these individual runs we find $0.332$, $0.259$, and $0.325$, respectively. Because the runs differ in duration and thermodynamic treatment, these numbers describe representative source variability rather than a matched-flow comparison. Figure~\ref{fig:grmhdlight} shows the recovered data, their temporal autocorrelations, and the nonoverlapping three-hour variability samples. The top-panel time axes use a nominal Sgr A* mass of $4.0\times10^6 M_\odot$ to express the recorded $10M$ sampling interval in hours.

\subsection{A controlled electron-prescription check}

\begin{figure*}[t]
 \centering
 \includegraphics[width=0.96\textwidth]{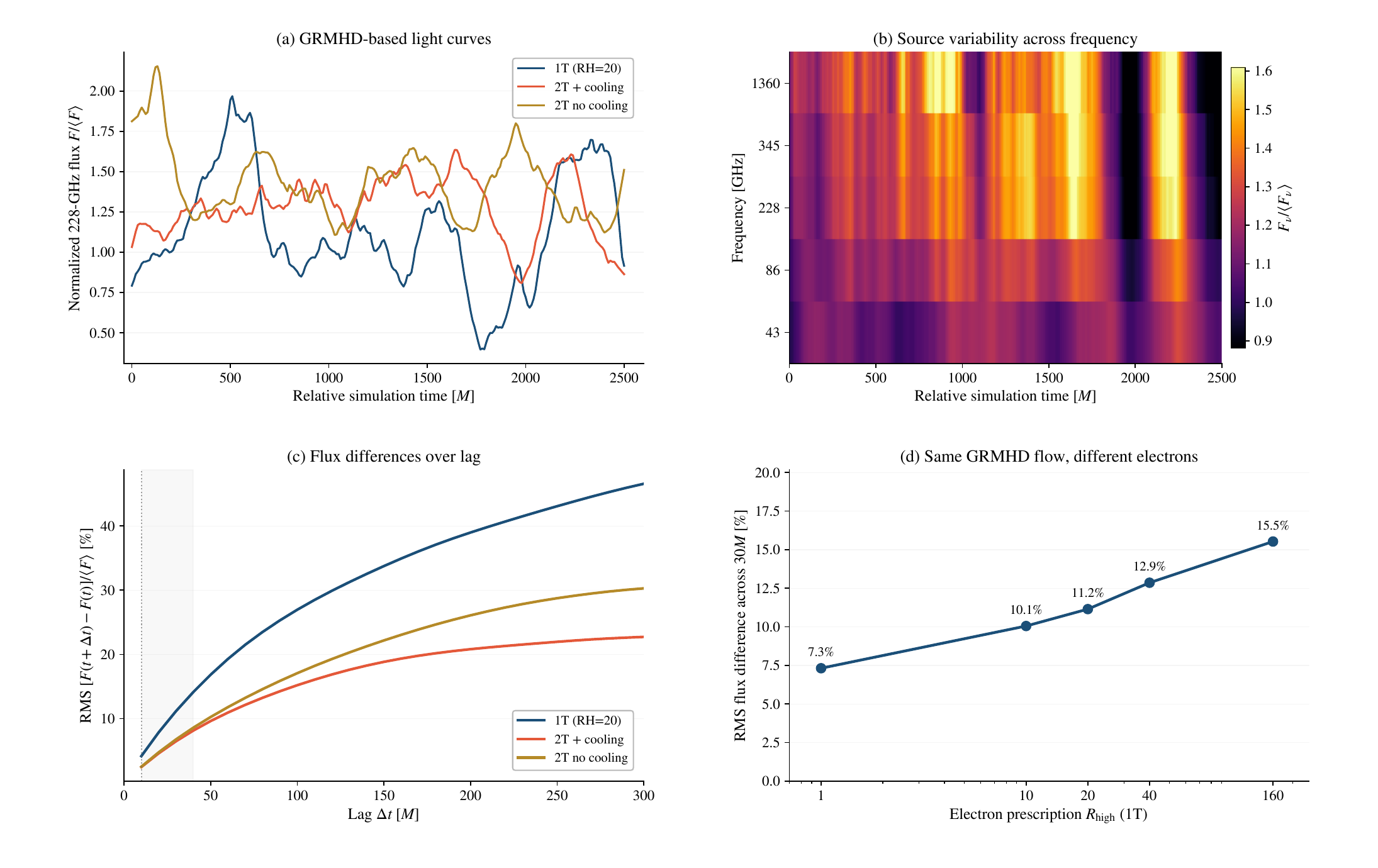}
 \caption{Our source-side analysis of the published Kerr GRMHD-based Sgr A* light curves of Ref.~\cite{Salas2025GRMHD} (public flux data in Ref.~\cite{Salas2025Data}). Panel (a): example 228-GHz normalized flux histories. Panel (b): time-frequency flux variations in one published simulation. Panel (c): the lag-dependent RMS flux increment $S_F$, as defined in Eq.~\eqref{eq:grmhdsf}. Panel (d): the $30M$ flux increment for five electron-temperature prescriptions applied to the same 1T simulated flow. The panels quantify how source emission changes with electron prescription and lag.}
 \label{fig:grmhdstructure}
\end{figure*}

For a lag $k\Delta t$ we define the dimensionless flux structure-function amplitude
\begin{equation}
 S_F(k\Delta t)=\frac{\big\langle[F(t+k\Delta t)-F(t)]^2\big\rangle^{1/2}}{\langle F\rangle}.
 \label{eq:grmhdsf}
\end{equation}
The released flux records have a native spacing $\Delta t=10M$, with $M=GM_\bullet/c^3$. We evaluate the structure function on the native sampled lags. At a lag of $30M$, we obtain $S_F=0.112$ (1T, $R_{\rm high}=20$), $0.0646$ (2T+C), and $0.0675$ (2T). For five electron-temperature prescriptions applied in post-processing to the \emph{same} 1T GRMHD flow, the corresponding values run from $0.0732$ ($R_{\rm high}=1$) to $0.1553$ ($R_{\rm high}=160$), a factor of $2.12$. These five results isolate the effect of the electron-temperature prescription on the same GRMHD flow. Figure~\ref{fig:grmhdstructure} also shows the time-lag dependence and a multi-frequency example.

\subsection{What the sampling resolves}

The median spike--Kerr order-delay differences in our M87*-scaled benchmark, 10.60 h and 3.36 h, correspond to $1.192M$ and $0.378M$ for $M_\bullet=6.5\times10^9 M_\odot$. Both shifts are shorter than the native $10M$ sampling interval of the published Sgr A* flux arrays. This comparison tests time-scale resolution in the archival data, rather than the performance of a proposed M87* array. \ref{app:grmhdtimeaudit} provides a direct plot of the cadence comparison.

\begin{figure*}[t]
 \centering
 \includegraphics[width=0.98\textwidth]{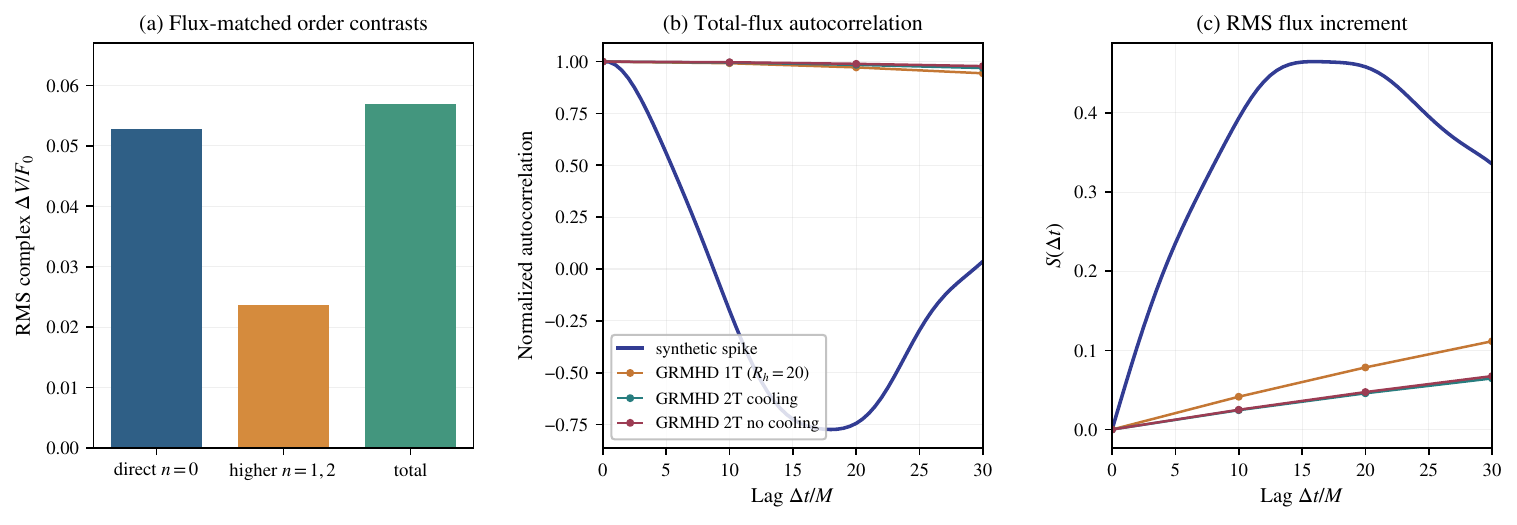}
 \caption{Archived flux-matched visibility contributions and an independent source-timescale check. Panel (a) shows complex RMS spike-minus-Kerr visibility differences for direct, higher-order, and total light on the stored 96-frame, 12-baseline shared-source track. They need not add linearly because the differences are complex. Panels (b,c) compare the mean-subtracted total-flux autocorrelation and normalized RMS flux increment for this semianalytic source with published 228-GHz Kerr GRMHD-based Sgr A* fluxes \cite{Salas2025GRMHD,Salas2025Data}. The sampling intervals are $0.5M$ for the synthetic source and $10M$ for the published GRMHD data.}
 \label{fig:sourcecalibration}
\end{figure*}

The external light curves provide an independent example of source fluctuations at comparable dimensionless lags. The published GRMHD flux autocorrelations decline more slowly than the archived semianalytic source autocorrelation; the 96-frame synthetic movie has relatively few overlapping samples at long lags. An explicit matched-lag comparison of the archived fluxes (Fig.~\ref{fig:sourcecalibration}) finds $S_F(30M)=0.335$ for the $96$-frame synthetic spike source, compared with $0.112$ for the published 1T flow and $0.0646$ and $0.0675$ for the two 2T prescriptions. The comparison shows how emission prescriptions and time scales can affect source-variability statistics. A direct GRMHD comparison of the two geometries will require compatible time-dependent images or fluid snapshots, with the source prescriptions and ray tracing held under common control.

\section{Discussion}\label{sec:discussion}

The main distinction is between finding a geometrical signal and assigning it to DM. Ring matching removes the leading size shift, while a smaller angular and time-domain deformation remains. The shared-source calculation shows that this remaining pattern survives the simulated station corruptions.

For the specified semianalytic population, separation becomes much less reliable when the source realization changes. The cross-array score correlations point to a multivariate degeneracy associated with the source realization. It also explains why our result is compatible with positive Kerr echo studies \cite{Wong2024Echo,Bezdekova2026,Zhang2025Echo}. Those studies ask whether strong-lensing structure is present. Here both hypotheses contain strong lensing, and the question is whether the smaller DM-minus-Kerr difference can be separated from an unknown source.

Cadence adds a second limitation. The broad echo window can be covered while a smaller geometry-dependent lag falls into the same daily gap. For a differential environmental test, source information and temporal phase coverage matter together.

The external GRMHD light curves in Sec.~\ref{sec:externalgrmhd} strengthen the source-side picture by showing how emission physics changes variability in a simulated accretion flow. They motivate geometry-matched GRMHD experiments and simultaneous source monitoring. A future Sgr A* application would also include scattering and rapid variability \cite{Johnson2018,Lu2016,EHT2022SgrAII,EHT2022SgrAIII}.

Higher sensitivity and longer baselines remain valuable because photon-ring information lives at high spatial frequency \cite{Johnson2020Universal,CardenasLupsasca2023,Jia2024,BHEX2024,GutierrezLara2026}. This motivates combining sensitive Fourier measurements with source monitoring and suitable delay-phase sampling.

The additional static-visibility check in Sec.~\ref{sec:interference} shows that direct and higher-order changes can partly cancel in the complex Fourier data. This is another reason why total-visibility contrast alone does not isolate a photon-ring deformation.

The high-inclination benchmark makes the geometric response particularly clear. The leading-ring matched Kerr comparison and independent-source tests quantify distinct parts of the problem: transport in a defined background and its robustness under changes in emission.

\ref{app:validation} records the validation checks, while \ref{app:roles} summarizes the role of each geometry family.

\section{Conclusions}\label{sec:conclusion}

After matching the leading Kerr ring scale, a strong spike-like deformation leaves distinct higher-order transport in the controlled effective-metric model. The effect appears in the critical parameters, the finite-order delays, and the slow-light visibility response. With a shared source, the same deformation remains visible after the simulated station corruptions.

The harder step is attribution. Fresh-source validation shows a large overlap between the Kerr and spike score distributions. The conditional score-scatter and sampling diagnostics highlight two practical concerns: source uncertainty and incomplete coverage of geometry-dependent delay phases. The resulting diagnostic scales provide concrete guidance for the controlled setups studied here.

Observing a photon echo and attributing a small departure from Kerr are distinct problems. Our geometric, radiative, visibility, and source-population tests make that distinction quantitative. They point toward joint higher-order timing, source monitoring, and improved delay-phase coverage as the most useful next steps.

\section*{Acknowledgments}
This work was supported by the Agencia Nacional de Investigaci\'on y Desarrollo (ANID), Chile, through FONDECYT Postdoctorado Grant No. 3260029.

\section*{Data, software and code availability}

The synthetic source-population scores, statistical checks, and scripts used for the present results are preserved in the project archive. Supporting numerical products and code are available from the author upon reasonable request. The published Sgr A* GRMHD-based flux products used in Sec.~\ref{sec:externalgrmhd} are publicly available at \href{https://doi.org/10.5281/zenodo.14793884}{Zenodo, doi:10.5281/zenodo.14793884} \cite{Salas2025Data}; the derived analysis scripts and numerical summaries are available from the author upon reasonable request.

\appendix
% Keep appendix equation hyperlink destinations distinct from main-text equations.
\renewcommand*{\theHequation}{appendix.\Alph{section}.\arabic{equation}}

\section{Metric checks for the rotating-spike benchmark}\label{app:yueaudit}

The photon calculation uses the effective rotating geometry of Ref.~\cite{Yue2026}, with $f=2qr$, $B=2mr$, the mass function in Eq.~\eqref{eq:massfunction}, and $q'=(m-q)/(2m-r)$. For $a/M_B=0.8$, the Hamilton--Jacobi separation residual is at most $7.1\times10^{-15}$ and the $q$--$m$ consistency residual is at most $2.2\times10^{-16}$. Removing the halo gives the Kerr horizon at $r_+/M_B=1.6$. These values verify the numerical consistency of the photon-geodesic implementation and its Kerr limit.

Our direct checks cover the metric relations, null-geodesic implementation, and static-seed field equations; the photon observables follow from the specified effective metric.

\section{Static-seed Einstein-equation check}\label{app:staticseed}

A useful independent check concerns the nonrotating limit of the rotating-spike benchmark. With $a=0$, its line element takes the form
\begin{equation}
 ds^2=-\left(1-\frac{2q}{r}\right)dt^2
 +\frac{dr^2}{1-2m/r}+r^2d\Omega^2.
\end{equation}
Writing $g_{tt}=-e^{2\nu}$, the spherical Einstein equations give
\begin{equation}
 8\pi\rho=\frac{2m'}{r^2},\qquad
 8\pi p_r=-\frac{2m}{r^3}
 +\frac{2}{r}\left(1-\frac{2m}{r}\right)\nu'.
\end{equation}
The defining relation $q'=(m-q)/(2m-r)$ implies
\begin{equation}
 \nu'=\frac{m}{r(r-2m)},\qquad p_r=0.
\end{equation}
The remaining spherical field equation, or equivalently conservation of this static anisotropic source, then gives
\begin{equation}
 p_{\perp}=\frac{\rho m}{2(r-2m)}.
\label{eq:staticpt}
\end{equation}
These relations characterize the effective anisotropic stress of the static seed.

For the baseline mass function in Eq.~\eqref{eq:massfunction}, $m-M_B$ and $m'$ both vanish at the onset $r_{\rm ISCO}$, so this piecewise static profile does not introduce a delta-function surface layer there. As a numerical check, we integrated the $q$ equation at the same baseline parameters and evaluated the two radial Einstein equations at 3000 points over $2.908\leq r/M_B\leq100$. In units $M_B=1$, the maximum absolute radial-pressure residual was $5.6\times10^{-19}$, the relative residual in $\nu'$ was below $1.1\times10^{-15}$, and the sampled ratio $p_{\perp}/\rho$ remained below $0.551$. The numerical residuals confirm the static-seed equations at the stated parameters.

\section{Image-order check after flux normalization}\label{app:ordercheck}

We use the saved 230-GHz target and Kerr movies to separate the direct image from higher orders. The complex visibilities satisfy $V=V_0+V_1+V_2$ in both models. We compare their stored tracks at baseline lengths $|\bm u|\leq8.5\,\mathrm{G}\lambda$ (946 of 1152 time--baseline samples), using the same source realization. The broader all-baseline diagnostic in the archived numerical audit uses all 1152 samples and therefore gives slightly different RMS values. We also allow one extra flux factor for the target: $g=\langle F_{K}\rangle/\langle F_{\rm sp}\rangle=1.2776$. This gives both models the same time-mean total flux.

\begin{figure*}[t]
 \centering
 \includegraphics[width=0.90\textwidth]{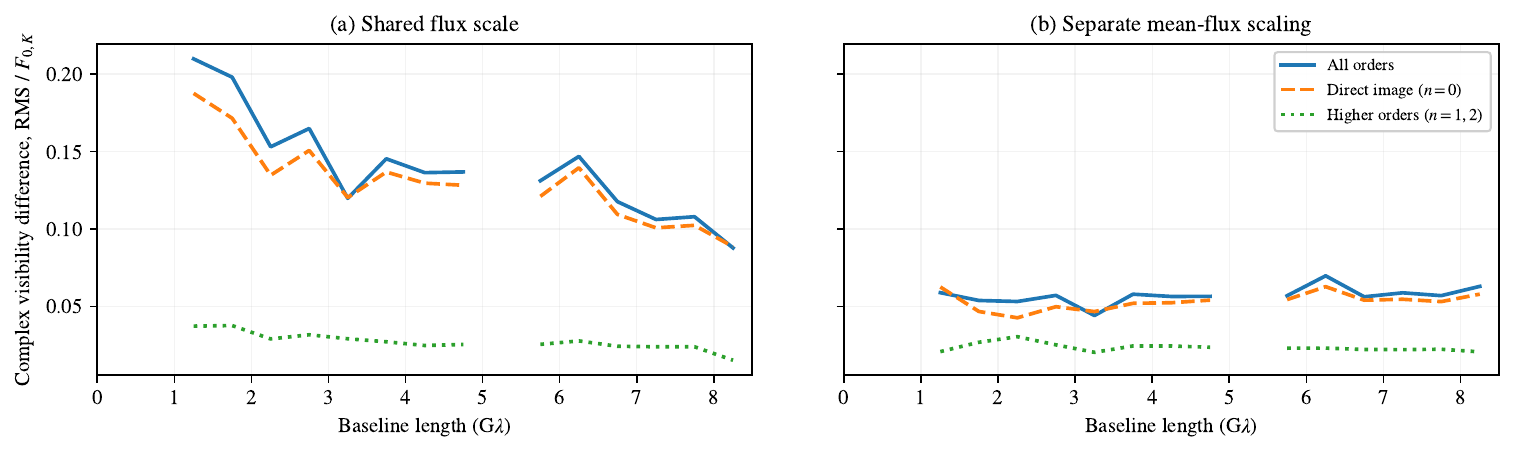}
 \caption{Complex target-minus-Kerr visibility differences from the archived 230-GHz movie, separated by image order. Panel (a) keeps the original common flux scale; panel (b) independently matches time-mean flux between geometries. Curves are RMS amplitudes in baseline-length bins on stored tracks up to $8.5\,\mathrm{G}\lambda$. The curves show the RMS magnitude of each complex-visibility contribution.}
 \label{fig:ordernorm}
\end{figure*}

Before this rescaling, the root-mean-square complex visibility difference is $0.1380F_{0,K}$. It falls to $0.0569F_{0,K}$ afterward. For the rescaled difference, the direct image contributes an RMS of $0.0528F_{0,K}$, while the combined $n=1,2$ contribution is $0.0237F_{0,K}$. These are complex amplitudes; the order components can interfere and their RMS values cannot simply be added. Figure~\ref{fig:ordernorm} shows their baseline dependence. The direct image makes a substantial contribution to the complex visibility difference. In the archived movie, both ray transport and source-dependent weighting contribute.

\section{Controlled inference metrics}\label{app:metrics}

For the population tests, the false-positive rate is measured relative to the array-specific 95th percentile of held-out Kerr scores. Power is the fraction of target realizations above this fixed threshold. Balanced accuracy is
\begin{equation}
 \mathrm{BA}=\frac{1}{2}(\mathrm{TPR}+\mathrm{TNR}),
\end{equation}
and ROC area is used as a threshold-independent ranking measure. The predefined gates were
\begin{equation}
\begin{aligned}
 \mathrm{FPR}&\le0.10, & \mathrm{power}&\ge0.75,\\
 \mathrm{AUC}&\ge0.90, & \mathrm{BA}&\ge0.85.
\end{aligned}
\end{equation}
The raw population likelihood also had a source-library stability gate. The driver-conditioned tests had a minimum-response-coverage gate.

Table~\ref{tab:popsizes} gives the sample sizes and uncertainty for the source-robust tests. The AUC intervals use class-stratified bootstrap resampling with 20,000 draws. Power intervals are Wilson 95\% intervals. The raw-closure rows use 32 training realizations per geometry and 24 held-out realizations per class. The two driver tests use separate development, validation, and sealed confirmatory designs.

\begin{strip}
\begin{minipage}{\textwidth}
\refstepcounter{table}\label{tab:popsizes}
\noindent\textbf{Table~\thetable.} Source-population sizes and uncertainty. Counts in the validation column are evaluable Kerr+spike realizations; parentheses give the full designed validation count when a predefined coverage failure is present. The AUC and power intervals are 95\% intervals.\par\smallskip
\centering
\small
\begin{tabular}{lllll}
\toprule
Test & Training/development & Validation & AUC [95\%] & Power [95\%] \\
\midrule
Raw closures, EHT-like & 32/geometry & $24+24$ & $0.658\ [0.497,0.807]$ & $0.083\ [0.023,0.258]$ \\
Raw closures, ngEHT-like & 32/geometry & $24+24$ & $0.682\ [0.526,0.830]$ & $0.125\ [0.043,0.310]$ \\
Raw closures, space-like & 32/geometry & $24+24$ & $0.623\ [0.458,0.776]$ & $0.083\ [0.023,0.258]$ \\
Scalar driver & $12+12$ & $15+15\ (16+16)$ & $0.507\ [0.293,0.716]$ & $0.000\ [0,0.204]$ \\
Spatial driver & $16+16$ & $20+19\ (20+20)$ & $0.500\ [0.313,0.692]$ & $0.0526\ [0.009,0.246]$ \\
\bottomrule
\end{tabular}
\end{minipage}
\end{strip}

The two driver designs use different fixed statistics, and their uncertainty intervals are evaluated separately.

Table~\ref{tab:nuisancecorr} reports the EHT-like Spearman correlations between the raw-closure score and the six source-design coordinates, evaluated separately for Kerr and spike.
\begin{table}[!t]
\caption{EHT-like Spearman $\rho_S$ correlations of the score with source parameters (24 held-out sources per class).}
\label{tab:nuisancecorr}
\centering\small
\begin{tabular}{lrr}
\toprule
Source coordinate & Kerr & Spike \\
\midrule
log-density amplitude & $-0.138$ & $+0.299$ \\
correlation time & $+0.357$ & $-0.314$ \\
reference optical depth & $+0.123$ & $-0.160$ \\
radial correlation length & $-0.197$ & $+0.190$ \\
azimuthal correlation length & $-0.157$ & $-0.076$ \\
rotation fraction & $+0.219$ & $+0.168$ \\
\bottomrule
\end{tabular}
\end{table}

The Gaussian $d$ conversion in Sec.~\ref{sec:req} is only a diagnostic for the chosen threshold and power target. The empirical score distributions remain the main result. Equation~\eqref{eq:scatterreq} combines source and instrumental scatter in the observed score.

\section{Temporal sampling of the external GRMHD check}\label{app:grmhdtimeaudit}

\begin{figure*}[t]
 \centering
 \includegraphics[width=0.80\textwidth]{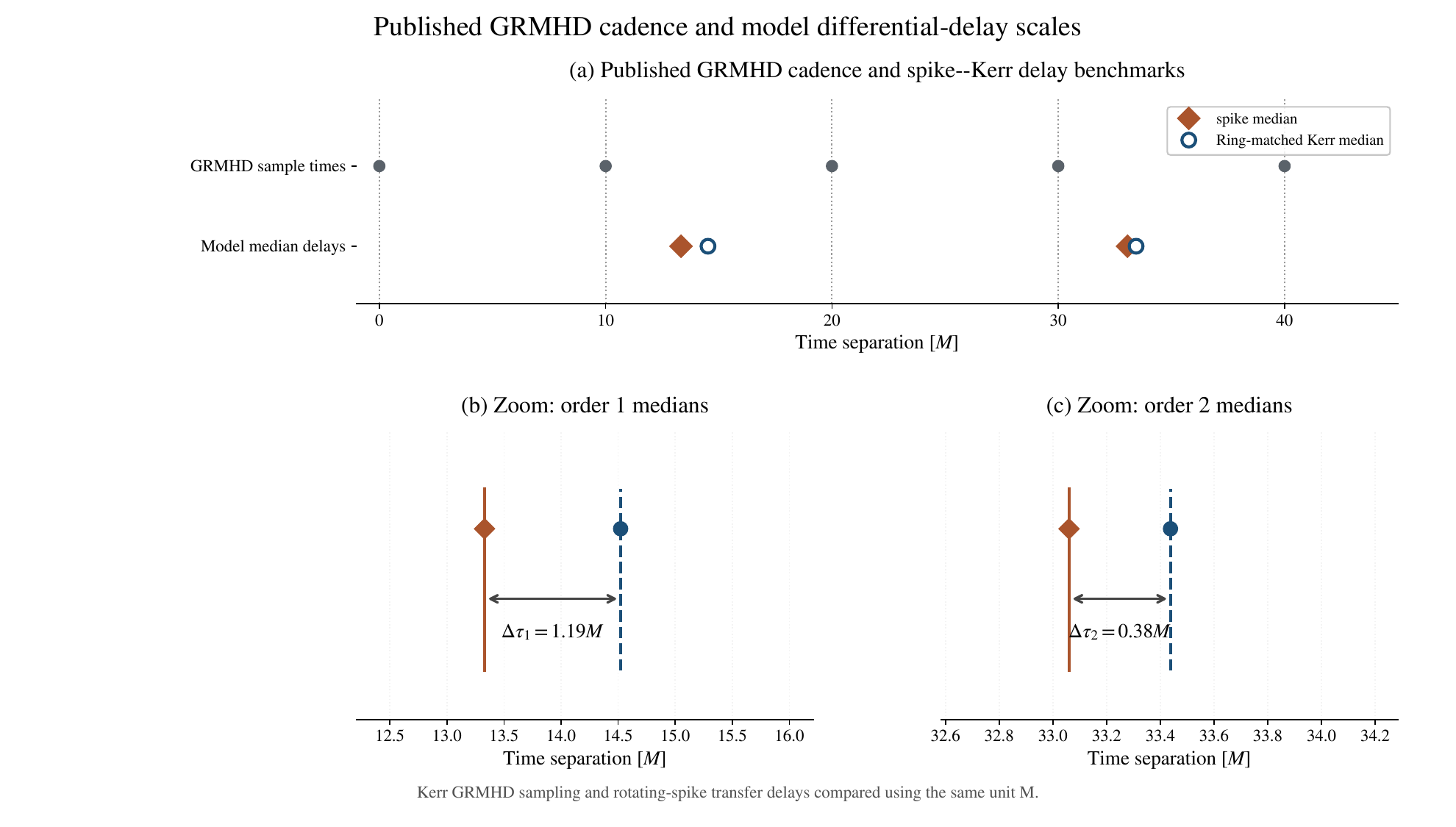}
 \caption{Sampling-scale audit. The published Sgr A* GRMHD-based light curves have a native cadence of $10M$. The spike--matched-Kerr differential median delays of $1.192M$ and $0.378M$ come from the M87*-scaled photon-transfer benchmark. The published Kerr GRMHD cadence and these transfer delays are shown together in gravitational time units.}
 \label{fig:grmhdtimeaudit}
\end{figure*}

Figure~\ref{fig:grmhdtimeaudit} places the native spacing of the published Kerr GRMHD-based flux samples beside the two differential delay shifts from our separate M87*-scaled DM model. The plot compares native sample intervals with the characteristic delays of the benchmark.

\section{Validation hierarchy and remaining checks}\label{app:validation}

We applied fixed validation checks throughout the calculation. The geometry stage checks spherical-orbit conditions, separability, the Kerr limit, and matched-ring residuals. The finite-order stage checks adaptive lensing-band convergence and keeps the exact separatrix out of the finite image-plane weight. The radiative stage compares full fine and coarse adaptive trees in complex visibility. The array stage repeats this check on the time-dependent station tracks. The source-population stages use separate development, validation, and reserved confirmatory populations with disjoint seeds and nuisance vectors.

The source-robust analysis stops at validation. When the validation criteria were not met, the reserved confirmatory population remained sealed and no further model tuning was performed.

All 36 fine/coarse visibility comparisons along the time-dependent station tracks satisfy the predefined convergence gate (Sec.~\ref{sec:oracle}). Two additional source-weighted matched-Kerr cases await refinement; an earlier Hernquist template-recovery test also remained below its response-coverage gate. The checked numerical implementation uses the project pipeline.

\section{Model roles and scope}\label{app:roles}

Table~\ref{tab:roles} lists the role of each geometry class. The strong spike is the primary benchmark, while the smoother models show how the problem changes for weaker near-critical deformations. The full source-robust analysis uses the strong spike, while the other profiles show the relative transport scale.
\begin{strip}
\begin{minipage}{\textwidth}
\refstepcounter{table}\label{tab:roles}
\noindent\textbf{Table~\thetable.} Role of each geometry class in the analysis. The last column identifies the main restriction when comparing model classes.\par\smallskip
\centering
\begin{tabular}{llll}
\toprule
Class & Role & Layer & Main caution \\
\midrule
Kerr & control & all & echoes also in Kerr \\
Einasto & smooth halo & geometry & scale-degenerate \\
cNFW & cored halo & geometry & convention dependent \\
Hernquist & cusp control & transfer & difficult low signal \\
Central spike & primary benchmark & full & $70^\circ$ benchmark; $17^\circ$ low-$i$ check \\
Solitonic & reserved & none & rotating metric required \\
\bottomrule
\end{tabular}
\end{minipage}
\end{strip}
%

% removed barrier test
% Use the physics bibliography style when supplied with the official EPJC template.
\renewcommand{\bibfont}{\small}
\bibliographystyle{spphys}
\bibliography{dark_echo}

\end{document}